\documentclass[11pt, reqno]{amsart}
\usepackage{graphicx,amsmath,amssymb,epsfig,float}
\usepackage{amsfonts, lscape}
\usepackage{import}
\usepackage{geometry}
\usepackage{verbatim}
\usepackage[section]{placeins}
\usepackage{float}
\usepackage[skip=2pt,font=footnotesize]{caption}
\usepackage{subcaption}
\usepackage{hyperref}
\hypersetup{
    colorlinks=true,
    urlcolor=blue,
    linkcolor=blue,
    citecolor=blue
}

\usepackage[
backend=biber,
style=apa,
sorting=nyt
]{biblatex}
\DeclareSourcemap{
  \maps[datatype=bibtex]{
    \map{
      \step[fieldset=issn, null]
      \step[fieldset=doi, null]
      \step[fieldset=url, null]
      \step[fieldset=urldate, null]
    }
  }
}

\restylefloat{table}
\theoremstyle{plain}

\newtheorem{assumption}{Assumption}

\newtheorem{definition}{Definition}

\newtheorem*{thA}{Theorem A}
\newtheorem*{thB}{Theorem B}

\numberwithin{equation}{section}
\theoremstyle{definition}

\graphicspath{ {FigsEtc/} }
\input{tcilatex}

\begin{document}
\title[Like Attract Like?]{\textbf{Like Attract Like?}\\
{\ \textbf{A Structural Comparison of Homogamy across Same-Sex and
Different-Sex Households }}}
\author[Ciscato]{Edoardo Ciscato$^{\flat }$}
\author[Galichon]{Alfred Galichon$^{\dag }$ }
\author[Gouss\'{e}]{Marion Gouss\'{e}{\small $^{\S }$}}
\date{First circulated version: November 21, 2014. This version: August 2018.
Galichon's research has been supported by NSF grant \# DMS-1716489, as well
as ERC grants FP7-295298, FP7-312503, FP7-337665. The authors would like to
thank the editor, James Heckman, and four anonymous referees as well as
Arnaud Dupuy, Sonia Oreffice, Bernard Salani\'{e} and Simon Weber for useful discussion. Accepted for publication by the \textit{Journal of Political Economy}
Volume 128, Number 2, February 2020, URL: \url{https://doi.org/10.1086/704611}.}

\begin{abstract}
In this paper, we extend Gary Becker's empirical analysis of the marriage
market to same-sex couples. Becker's theory rationalizes the well-known
phenomenon of homogamy among different-sex couples: individuals mate with
their likes because many characteristics, such as education, consumption
behaviour, desire to nurture children, religion, etc., exhibit strong
complementarities in the household production function. However, because of
asymmetries in the distributions of male and female characteristics, men and
women may need to marry \textquotedblleft up\textquotedblright\ or
\textquotedblleft down\textquotedblright\ according to the relative shortage
of their characteristics among the populations of men and women. Yet, among
same-sex couples, this limitation does not exist as partners are drawn
from the same population, and thus the theory of assortative mating would
boldly predict that individuals will choose a partner with nearly identical
characteristics. Empirical evidence suggests a very different picture: a
robust stylized fact is that the correlation of the characteristics is in
fact weaker among same-sex couples. In this paper, we build an equilibrium
model of same-sex marriage market which allows for straightforward
identification of the gains to marriage. We estimate the model with
2008-2012 ACS data on California and show that positive assortative mating
is weaker for homosexuals than for heterosexuals with respect to age and
race. Our results suggest that positive assortative mating with respect to
education is stronger among lesbians, and not significantly different when
comparing gay men and married different-sex couples. As regards labor market
outcomes, such as hourly wages and working hours, we find some indications
that the process of specialization within the household mainly applies to
different-sex couples.
\end{abstract}
\maketitle
\setcounter{page}{1}

{\footnotesize \noindent \textbf{Keywords}: sorting, matching, marriage
market, homogamy, same-sex households, roommate problem.\newline
}




\newpage

\section{Introduction}
\setcounter{equation}{0}
How individuals sort themselves into marriage has important implications for
income distribution, labor supply, and inequality \parencite{becker1973theory}.
Strong evidence shows that assortative mating in marriages accounts for a non-negligible part of income inequality across households %
\parencite{eika2014educational}.

Individuals tend to mate with their likes, a pattern called \emph{homogamy}. However, because of asymmetries
between the distributions of the characteristics in male and female
populations, homogamy cannot be perfect among different-sex couples. In other
words, heterosexuals cannot always find a \textquotedblleft clone%
\textquotedblright{} of the opposite sex to match with. A large body of the
literature has noticed that, up until recently, \textquotedblleft men
married down, women married up\textquotedblright\ due to the sex asymmetry
in educational achievement that has only recently started to fade %
\parencite{goldin2006homecoming}. Gender asymmetries exist in other dimensions
such as biological characteristics (windows of fertility\footnote{%
Women's fertility rapidly declines with age, whereas men's fertility does
not. Biologists and anthropologists argue that this dissymmetry could
explain the well-documented preference of men for younger women %
\parencite{hayes1995age,kenrick1992age}. \cite{low2013pricing} evaluates this
young age premium for women and names it \textquotedblleft reproductive
capital,\textquotedblright\ as it gives them an advantage on the marriage
market over older women.}, life expectancy, bio-metric characteristics),
psychological traits, economic attributes (due to the gender wage gap),
ethnic and racial characteristics
\parencite[immigration is not symmetric across
sexes, see][]{weiss2013hypergamy} or demographic characteristics (some
countries, such as China, have comparatively more imbalanced gender ratios).

Homogamy has been famously studied by Becker's seminal analysis of the
family. \cite{becker1973theory} expects most non-labor market traits, such
as \textquotedblleft intelligence, height, skin color, age, education,
family background or religion\textquotedblright{}, to be complements.
However, he also suggests that some attributes could be substitutes; in
particular, Becker suggests that we should observe a negative correlation
between some labor market traits such as wage rates because of household
specialization\footnote{\cite{chiappori2012fatter} model a Becker-like
marriage market with sorting on a unidimensional index. The estimation of
such index reveals that high values in some attributes can compensate for
poor values in others, thus showing that sorting is based on trade-offs
between traits.}. In order to provide a structural explanation of homogamy,
Becker proposed a model of positive assortative mating (PAM) in which men
and women are characterized by some socio-economic \textquotedblleft
ability\textquotedblright\ index. In this model, the marriage market clears
so that men are matched with women that are as close as possible to them in
terms of this index, which subsumes all the characteristics that matter on
the marriage market. The (strong) prediction of Becker's model is that the
rank of the husband's index in the men's population is the same as the
wife's in the women's population. However, this does not imply that the
partners' indices are identical: they would be so only if the distributions
of the indices were the same for both men and women's populations.

This analysis of the marriage market has attracted wide attention in the
economic literature, in spite of its shortcomings. One shortcoming is that
it originally refers to different-sex unions only. However, in a growing
number of countries, same-sex couples have gained legal recognition, and the
institutions of civil partnership and marriage no longer require that the
partners must be of opposite sex. This official recognition is the result of
several legal disputes and social activism by the gay and lesbian communities%
\footnote{%
Public actions for gay rights acknowledgment are often considered to
have started in 1969, in New York City. See \cite{eskridge1993history} and %
\cite{sullivan2009same} for a detailed history and a full overview of the
arguments in favor of and against same-sex marriage.}. The issue of whether
to recognize same-sex unions has long been a topical subject in many
countries, since it challenges the traditional model of the family. From
both an economic and a legal point of view, the definition of what
\textquotedblleft family\textquotedblright\ means has relevant political
implications as long as this term is present - and is generally central - in
many modern constitutions and legal systems. Consequently, family units
benefit from a special attention of policy-makers. Therefore, a discussion
of the issues related to same-sex marriage - remarkably at policy level -
requires a good understanding of the similarities and differences in the
household dynamics among same-sex and different-sex couples. Besides, it is
important to remember that the legal recognition of same-sex couples is only
one of many transformations that the institution of the family has gone
through in the last decades \parencite{stevenson2007marriage,stevenson2008}.
Finally, since more and more data on same-sex unions have been made
available, the extension of the economic analysis of family to the
gay and lesbian population can now be taken to data.

While it is natural to consider an extension of Becker's model to same-sex
households, it is worth noting that the previous considerations on
asymmetries between men's and women's distributions only hold as long as
each partner comes from a separate set according to his/her sex. On the
same-sex marriage market, the two partners are drawn from the same
population and the distributions of the characteristics are the same. Hence,
the assortative mating theory pushed to its limits implies that, in this
setting, partners should be exactly identical, i.e., each individual will
choose to marry someone with identical characteristics.

In spite of such theoretical predictions, facts suggest a very different
picture. Recent empirical results on the 1990 and 2000 American Census show
that same-sex couples have less correlated attributes than different-sex
ones, at least in terms of a variety of non-labor traits, including racial
and ethnic background, age and education %
\parencite{jepsen2002empirical,schwartz2009assortative}. Studies on Norway,
Sweden \parencite{andersson2006demographics} and Netherlands %
\parencite{verbakel2014assortative} led to similar findings. In order to explain
these systematic differences, the literature has suggested several possible
reasons. A first consideration is that gay people might be forced to pick
from a restricted pool because of their smaller numbers in the population,
thus having a narrower choice when selecting their partner, resulting in a
more diverse range of potential matches %
\parencite{harry1984,kurdek1987partner,andersson2006demographics,schwartz2009assortative,verbakel2014assortative}%
. Furthermore, gay men and lesbians have been found to be more likely to live in
urban neighborhoods than heterosexuals, and since diversities in
socio-economic traits are stronger in cities, this facilitates the crossing
of racial and social boundaries %
\parencite{black2002gay,rosenfeld2005independence,black2007economics}. In light
of these observations, one could argue that the same-sex marriage market
is faced with stronger search frictions. Nevertheless, this might not
necessarily be the case if the potential partners gather in specific
locations, as it happens in cities and neighborhoods that are considered
\textquotedblleft gay-friendly\textquotedblright\ \parencite{black2007economics}%
. Other analysts argue that gay people may have different preferences than
heterosexuals, as they tend to be less conservative than straight
individuals. Some explanations in this regard point out that, since
homosexuality is still considered in some cultures as at odds with
prevailing social norms, gay men and lesbians might grow less inclined to passively
accept social conventions, and consequently they would end up choosing their
partner with fewer concerns about his/her background traits\footnote{%
Note that household location choice and social norms are strictly related:
it has been reported that gay people often leave their town of origin and
escape social pressure exerted by relatives and acquaintances and go living
in larger cities reputed to be gay/lesbian-friendly %
\parencite{rosenfeld2005independence}. Analogously, they are aware that
they have more probabilities of avoiding discrimination by achieving higher
educational levels and orienting their professional choices toward congenial
working environments \parencite{blumstein1983american,verbakel2014assortative}.} %
\parencite{blumstein1983american,meier2009young,schwartz2009assortative}. The
detachment from the community of origin and the research for more tolerant
surroundings have an influence both on values and social norms and on the
heterogeneity of interpersonal ties.

A part of these explanations has to do with individual preferences, whereas
another part has to do with demographics, i.e., the distribution of the
characteristics in the population. It is clear that the explanations listed
in the former paragraph, while different in nature, are not mutually
exclusive, but all contribute to a better understanding of the equilibrium
patterns. For instance, a high correlation in education may arise from
individual tastes (as individuals could find more desirable to match with a
partner of similar educational background), but also from demographics
(indeed, if some educational category represents a large share of
individuals, this will increase odds of unions within this category, thus
mechanically increasing the correlation in education). When comparing the
heterosexual and homosexual population, this is particularly relevant as
their compositions present significant differences
\parencite[e.g.,
gay people are, on average, more educated than heterosexuals; see
][]{black2007economics}.

In this paper, we focus on differences in marital gains: we would like to
compare the structure of complementarity and substitutability across
same-sex and different-sex households. In order to do so, we need a
methodology that helps us interpret the observation of matching patterns and
disentangle the role played by household interactions from external
demographic factors. This is achieved through a structural approach, which
allows us to estimate the parameters of the marital surplus function in
order to fit the patterns actually observed. Hence, this approach will
require an equilibrium model of matching.

In the wake of \cite{becker1973theory,becker1981treatise}, the economic
literature has modeled the marriage market as a bipartite matching game with
transferable utility. A couple consists of two partners coming each from a
separate or identical subpopulation (respectively, in the case of
different-sex and same-sex unions). Both partners are characterized by
vectors of attributes, such as education, wealth, age, physical
attractiveness, etc. It is assumed that, when two partners with respective
attributes $x$ and $y$ form a pair, they generate a surplus equal to $\Phi
(x,y)$, which is shared endogenously between them. In the case of separate
subpopulations (different-sex marriage), the landmark contribution of %
\cite{choosiow2006} showed that the surplus function $\Phi $ can easily be
estimated based on matching patterns modulo a distributional assumption on
unobservable variations in preferences, and was followed by a rich
literature
\parencite[][to cite a
few]{fox2010identification,galichon2012cupid,chiappori2017partner}. %
\cite{dupuy2014personality} extended Choo and Siow's model to the case of
continuous attributes and propose the convenient bilinear parameterization $%
\Phi \left( x,y\right) =x^{\prime }Ay$, where $A$ is a matrix called
\textquotedblleft affinity matrix\textquotedblright\ whose terms reflect the
strength of assortativeness between two partners' attributes. However, the
bipartite assumption is restrictive and does not allow to estimate the
surplus on same-sex marriage markets, and, to the best of our knowledge, no
such estimation procedure is proposed in the literature. In a theoretical
paper, \cite{chiappori2012roommate} focus on stable matchings in a finite
population and show that, when the population to be matched is doubled by
cloning, the same-sex marriage problem, or \textquotedblleft unipartite
matching problem\textquotedblright\, can be mathematically reformulated as a
heterosexual matching problem, or \textquotedblleft bipartite matching
problem\textquotedblright\footnote{%
In another recent theoretical work, \cite{peski2017large} extends the NTU
framework of \cite{dagsvik2000aggregation} and \cite{menzel2013large} and
discusses the existence of stable matching in the unipartite case. %
\cite{fox2018estimating} proposes an empirically tractable TU framework
that generalizes both the bipartite and the unipartite case, and applies it
to the car parts industry.}. In section \ref{sec:model} of the present
paper, we apply an analogous reasoning to the empirically tractable,
large-population, two-sided matching model of \cite{dupuy2014personality},
in order to adapt their empirical strategy to the same-sex marriage market.

A few papers already deal with the issue of assortativeness among same-sex
households, although none of them allows to draw conclusions on the
structural parameters of the surplus function that drives the
assortativeness. The most relevant benchmarks for the empirical results of
this work are the aforementioned \cite{jepsen2002empirical} and %
\cite{schwartz2009assortative}. Both papers make use of the American census
data (1990 and 1990/2000 respectively) and find that members of
different-sex couples are more alike than those of same-sex ones with
respect to non-labor market traits. The heterogeneity in assortativeness is
measured in a logit framework containing dedicated parameters for homogamy.
In general, in a logit framework individuals choose their best option among
all possibilities. However, this fails to take into account the fact that
matching takes place under scarcity constraint on the various
characteristics. In the present paper, we estimate a model of matching in
which agents compete for a partner; our measures of assortativeness are
given by the parameters of the surplus function in each market (gay, lesbian
and heterosexual).

The contributions of the present paper are twofold. On a methodological
level, this paper is the first to propose a structural estimator of the
matching surplus which applies to same-sex households, or, more generally,
to instances of the unipartite matching problem. On an empirical level, we
provide evidence by means of a structural analysis that, as concerns age and
ethnicity, different-sex couples exhibit a higher degree of assortativeness
than same-sex ones. While we find, in line with previous results, that
sorting on education is stronger among lesbians with respect to different-sex
couples, our results suggest that assortativeness on education is not
significantly different when comparing gay male and married different-sex couples.
Further, we also look at labor market traits such as hourly wages and
working hours. Comparing assortativeness on labor market outcomes between
same-sex and different-sex couples hints to different family dynamics and
differences in the household specialization process.
Finally, we briefly discuss the estimates of the mutually exclusive affinity
indices obtained through our saliency analysis.

The rest of the paper is organized as follows. Section \ref{sec:model} will
present the model and section \ref{sec:estimation} the estimation procedure.
We describe our data in section \ref{sec:data} and our results in section %
\ref{sec:results}. Section \ref{sec:conclusion} concludes.

\section{The model\label{sec:model}}

In what follows, it is assumed that the full type of each individual, i.e.,
the complete set of all individual characteristics that matter for the
marriage market (physical attributes, psychological traits, socio-economic
variables, sex, sexual orientation, etc.), is fully observed by market
participants. Each individual is characterized by a vector of observable
characteristics $x\in \mathcal{X}=\mathbb{R}^{K}$, which constitutes his or
her observable type. However, following \cite{choosiow2006}, we allow for a
certain degree of unobserved heterogeneity by assuming that agents
experience variations in tastes that are not observable to the analyst, but
are observable to the agents. In this paper, types are assumed to be
continuous, as in \cite{dupuy2014personality}, hereafter DG, and %
\cite{menzel2013large}. Assume that the distribution of the characteristics
$x$ has a density function $f$ with respect to the Lebesgue measure. Without
loss of generality, the marginal distribution of the attributes is assumed
to be centered, i.e. $\mathbb{E}[X]=0$.

\subsection{Populations}

A \emph{pair} is an ordered set of individuals, denoted $\left[ x_{1},x_{2}%
\right] $ where $x_{1},x_{2}\in \mathcal{X}$, in which the order of the
partner matters, which implies that the pair $\left[ x_{1},x_{2}\right] $
will be distinguished from its inverse twin $\left[ x_{2},x_{1}\right] $. In
empirical datasets, $x_{1}$ will often be denominated \textquotedblleft head
of the household\textquotedblright\ and $x_{2}$ \textquotedblleft spouse of
the head of the household\textquotedblright \thinspace\ even though this
denomination is used mainly for practical reasons and cannot be fully
representative of the actual roles in the household\footnote{%
We will come back in section \ref{par:testSymmetry} to this assumption that
the roles of partners are exchangeable, which we test using a number of
proxies for asymmetric household roles.}. A \emph{couple} is an unordered
set of individuals $\left( x_{1},x_{2}\right) $, so that the couple $\left(
x_{1},x_{2}\right) $ coincides with the couple $\left( x_{2},x_{1}\right) $.
A \emph{matching} is the density of probability $\pi \left(
x_{1},x_{2}\right) $ of drawing a couple $\left( x_{1},x_{2}\right) $. Pairs
$\left[ x_{1},x_{2}\right] $ and $\left[ x_{2},x_{1}\right] $ stand for the
same couple, so that the density $\pi \left( x_{1},x_{2}\right) $ is the sum
of the density of $\left[ x_{1},x_{2}\right] $ and of the density of $\left[
x_{2},x_{1}\right] $, hence the symmetry condition $\pi \left(
x_{1},x_{2}\right) =\pi \left( x_{2},x_{1}\right) $ holds. This symmetry
constraint means that the position of the individual must not matter and
thus that there are no predetermined \textquotedblleft
roles\textquotedblright\ within the couple that would be relevant for the
analysis\footnote{\cite{candelon2015hierarchical} extend %
\cite{chiappori2012roommate}'s analysis to a model where agents form
couples with endogenously assigned roles according to their characteristics.
The model is applied to team formation in professional road cycling. %
\cite{fox2018estimating} employs a very general many-to-many matching
framework where agents self-select to be buyers or sellers upon a meeting.
In both cases, hierarchy (leader vs assistants) or roles (buyers vs sellers)
are clearly defined upon a match and observed in the data. This is unlikely
to be the case when it comes to more complex and long-lasting relationships
such as marriage.}.

We shall impose assumptions that will ensure that everyone is matched at
equilibrium, hence the density of probability of type $x\in \mathcal{X}$ in
the population is given by $\int_{\mathcal{X}}\pi (x,x^{\prime })dx^{\prime
} $, which counts the number of individuals of type $x$ matched either as
the head of household in a couple $\left[ x,x^{\prime }\right] $, or as the
spouse of the head in a couple $\left[ x^{\prime },x\right] $. Thus, we are
led to assume:

\begin{assumption}[Populations]
\label{ass:pop}The density $\pi \left( x,x^{\prime }\right) $ over couples
satisfies $\pi \in \mathcal{M}^{sym}\mathcal{(}f\mathcal{)}$, where%
\begin{equation*}
\mathcal{M}^{sym}\mathcal{(}f\mathcal{)}=\left\{ \pi \geq 0:%
\begin{pmatrix}
\int_{\mathcal{X}}\pi (x,x^{\prime })dx^{\prime }=f(x)~\forall x\in \mathcal{%
X} \\
\pi (x_{1},x_{2})=\pi (x_{2},x_{1})~\forall x_{1},x_{2}\in \mathcal{X}%
\end{pmatrix}%
\right\} .
\end{equation*}
\end{assumption}

In contrast, in the classical bipartite problem, we try to match optimally
two distinct populations (men and women) which are characterized by the same
space of observable variables $\mathcal{X}$, and it is assumed that the
distribution of the characteristics among the population of men has density $%
f$, while the density of the characteristics among the population of women
is $g$. In this setting, the set of feasible matchings is typically given
by:
\begin{equation*}
\mathcal{M(}f,g\mathcal{)}=\left\{ \pi \geq 0:%
\begin{pmatrix}
\int_{\mathcal{X}}\pi (x,y)dy=f(x)~\forall x\in \mathcal{X} \\
\int_{\mathcal{X}}\pi (x,y)dx=g\left( y\right) ~\forall y\in \mathcal{X}%
\end{pmatrix}%
\right\}
\end{equation*}

Hence, $\pi \in \mathcal{M}^{sym}\mathcal{(}f\mathcal{)}$ if and only if $%
\pi \in \mathcal{M(}f,f\mathcal{)}$ and $\pi \left( x_{1},x_{2}\right) =\pi
\left( x_{2},x_{1}\right) $. Thus the feasibility set in the unipartite
problem and in the bipartite problem differ only by the additional symmetry
constraint in the unipartite problem.

\subsection{Preferences}

We now model preferences. Following DG, it is assumed that a given
individual $x$ does not have access to the whole population, but only to a
set of acquaintances $\left\{ z_{k}^{x}:k\in \mathbb{Z}_{+}\right\} $,
randomly drawn, which is described below.

\begin{assumption}[Preferences]
\label{ass:prefs}An individual of type $x$ matched to an individual of type $%
x^{\prime }$ enjoys a surplus which is the sum of three terms:

(i) the systematic part of the pre-transfer matching surplus enjoyed by $x$
from his/her match with $x^{\prime }$, denoted $\alpha \left( x,x^{\prime
}\right) $.

(ii)\ an endogenous utility transfer from $x^{\prime }$ to $x$, denoted $%
\tau \left( x,x^{\prime }\right) $. This quantity can be either positive or
negative; we assume utility is fully transferable, hence feasibility imposes
$\tau \left( x,x^{\prime }\right) +\tau \left( x^{\prime },x\right) =0$.

(iii) a \textquotedblleft sympathy shock\textquotedblright\ $\left( \sigma
/2\right) \varepsilon ^{x}$, which is stochastic conditional on $x$ and $%
x^{\prime }$, and whose value is $-\infty $ if $x$ is not acquainted with an
individual $x^{\prime }$. The quantity $\sigma /2$ is simply a scaling
factor. More precisely, the set of acquaintances is an infinite countable
random subset of $\mathcal{X}$; it is such that $(z_{k}^{x},\varepsilon
_{k}^{x})$ are the points of a Poisson process on $\mathcal{X}\times \mathbb{%
R}$ of intensity $dz\times e^{-\varepsilon }d\varepsilon $.
\end{assumption}

\bigskip

While the stochastic structure of the unobserved variation in preference
described in part (iii) of Assumption~\ref{ass:prefs} may appear complex, it
is in fact a very natural extension of the logit framework to the continuous
case, as we now argue. Indeed, it will imply that the individual
maximization program of an agent of type $x$ with this set of acquaintances
is%
\begin{equation}
\max_{k\in \mathbb{Z}_{+}}\alpha \left( x,z_{k}^{x}\right) +\tau \left(
x,z_{k}^{x}\right) +\frac{\sigma }{2}\varepsilon _{k}^{x},  \label{indivMax}
\end{equation}%
where the utility of matching with acquaintance $k$ yields a total surplus
which is the sum of three terms, the systematic pre-transfer surplus, the
transfer, and the sympathy shock. Define the systematic quantity of surplus
at equilibrium $U$ by
\begin{equation*}
U\left( x,x^{\prime }\right) :=\alpha \left( x,x^{\prime }\right) +\tau
\left( x,x^{\prime }\right)
\end{equation*}%
thus an individual of type $x$ maximizes $U\left( x,z_{k}^{x}\right) +\left(
\sigma /2\right) \varepsilon _{k}^{x}$ over the set of his/her
acquaintances, which are indexed by $k$. This induces an aggregate demand
over the type space. Indeed, it follows from the continuous logit theory
initiated in \cite{dagsvik1994discrete} that the conditional probability
density of an individual of type $x$ matching with a partner of type $%
x^{\prime }$ is%
\begin{equation}
\pi (x^{\prime }|x)=\frac{\exp \frac{U(x,x^{\prime })}{\sigma /2}}{\int_{%
\mathcal{X}}\exp \dfrac{U(x,x^{\prime })}{\sigma /2}dx^{\prime }}.
\label{eq:1}
\end{equation}%
It is clear from expression \eqref{eq:1} that this is a generalization of
the logit framework to the continuous case.

\bigskip

Note that, by the property of independence of irrelevant alternatives (IIA)\
of the logit model, we do not need to describe the utilities of unmatched
agents as long as the distributions of their stochastic parts are assumed to
remain in the logit setting. Indeed, in the dataset we use, all agents are
matched. Of course, one may worry about a potential equilibrium selection
issue, i.e., that being matched affects the distributions of the
agents' unobserved heterogeneity; however, in the logit setting, the IIA
property guarantees that the distributions are preserved even after the
selection, as shown in appendix D of DG. This is
the reason why we consider a model where everyone is matched at equilibrium.

\subsection{Equilibrium}

Next, we define equilibrium in this framework. Denote
\begin{equation*}
\Phi \left( x,x^{\prime }\right) :=\alpha \left( x,x^{\prime }\right)
+\alpha \left( x^{\prime },x\right) =U\left( x,x^{\prime }\right) +U\left(
x^{\prime },x\right)
\end{equation*}%
the systematic part of the joint surplus\footnote{%
Note that $\Phi $ is symmetric by definition, but $\alpha $ has no reason to
be symmetric. Mathematically speaking, $\Phi $ is (twice) the symmetric part
of $\alpha $.} between $x$ and $x^{\prime }$. It follows from \eqref{eq:1}
and symmetry of $\pi $ that%
\begin{eqnarray}
\left( \sigma /2\right) \ln \pi \left( x,x^{\prime }\right) &=&U(x,x^{\prime
})-a\left( x\right) =U(x^{\prime },x)-a\left( x^{\prime }\right) ,
\label{intermed} \\
\text{where }a(x) &:&=\dfrac{\sigma }{2}\log \int_{\mathcal{X}}\frac{1}{%
f\left( x\right) }\exp \dfrac{U(x,x^{\prime })}{\sigma /2}dx^{\prime }.
\label{expr_a}
\end{eqnarray}

Substituting out for $U$ in \eqref{intermed} yields the following equation,
which expresses optimality in individual decisions:%
\begin{equation}
\log \pi \left( x,x^{\prime }\right) =\dfrac{\Phi \left( x,x^{\prime
}\right) -a(x)-a(x^{\prime })}{\sigma },  \label{eqSymSchrodinger}
\end{equation}

At equilibrium, the value of $a\left( .\right) $ is determined by
market-clearing condition $\int_{\mathcal{X}}\pi \left( x,x^{\prime }\right)
dx^{\prime }=f\left( x\right) $, that is%
\begin{equation}
\int_{\mathcal{X}}\exp \left( \dfrac{\Phi \left( x,x^{\prime }\right)
-a(x)-a(x^{\prime })}{\sigma }\right) dx^{\prime }=f\left( x\right) .
\label{mktClearing}
\end{equation}

We can now define our equilibrium matching concept.

\begin{definition}
The density $\pi $ is an equilibrium matching if and only if there is a
function $a\left( .\right) $ such that both optimality equations %
\eqref{eqSymSchrodinger} and market clearing equations \eqref{mktClearing}
are satisfied.
\end{definition}

\bigskip

The main results on equilibrium characterization are summarized in the
following statement:

\begin{thA}
Under Assumptions \eqref{ass:pop} and \eqref{ass:prefs}:

(i) The equilibrium matching $\pi \left( x,x^{\prime }\right) $ is the
unique solution to%
\begin{equation}
\underset{\pi \in \mathcal{M(}f,f\mathcal{)}}{\text{max}}\iint_{\mathcal{%
X\times X}}\Phi (x,x^{\prime })\pi (x,x^{\prime })dxdx^{\prime }-\sigma
\mathcal{E}\left( \pi \right) ,  \label{opt}
\end{equation}%
where $\mathcal{E}\left( \pi \right) $ is defined by
\begin{equation}
\mathcal{E}\left( \pi \right) =\iint_{\mathcal{X\times X}}\pi (x,x^{\prime
})\ln \pi (x,x^{\prime })dxdx^{\prime }.  \label{defineEntropy}
\end{equation}

(ii) The expression of $\pi \left( x,x^{\prime }\right) $ is given by%
\begin{equation}
\pi \left( x,x^{\prime }\right) =\exp \left( \dfrac{\Phi \left( x,x^{\prime
}\right) -a(x)-a(x^{\prime })}{\sigma }\right) ,  \label{exprPi}
\end{equation}%
where $a\left( .\right) $ is a fixed point of $F$, which is given by%
\begin{equation}
F\left[ a\right] (x)=\sigma \log \int_{\mathcal{X}}\exp \left( \dfrac{\Phi
\left( x,x^{\prime }\right) -a(x^{\prime })}{\sigma }\right) dx^{\prime
}-\sigma \log f\left( x\right) .  \label{operator}
\end{equation}
\end{thA}

\begin{proof}
By DG, Theorem 1, Problem \eqref{opt} has a unique solution which can be
expressed as
\begin{equation*}
\pi \left( x,x^{\prime }\right) =\exp \left( \frac{\Phi \left( x,x^{\prime
}\right) -a\left( x\right) -b\left( x^{\prime }\right) }{\sigma }\right)
\end{equation*}%
for some $a\left( x\right) $ and $b\left( x^{\prime }\right) $ determined by
$\pi \in \mathcal{M}\left( f,f\right) $. By the symmetry of $\Phi $ and by
the symmetry of the constraints implied by $\pi \in \mathcal{M}\left(
f,f\right) $, then $\tilde{\pi}\left( x^{\prime },x\right) :=\pi \left(
x,x^{\prime }\right) $ is also solution to \eqref{opt}. By uniqueness, $%
\tilde{\pi}=\pi $, thus $\pi \left( x,x^{\prime }\right) =\pi \left(
x^{\prime },x\right) $. As a result, $b\left( x\right) =a\left( x\right) $,
where $a$ is determined by%
\begin{equation*}
\int \exp \left( \frac{\Phi \left( x,x^{\prime }\right) -a\left( x\right)
-a\left( x^{\prime }\right) }{\sigma }\right) dx^{\prime }=f\left( x\right)
\end{equation*}%
QED.
\end{proof}

\bigskip

This result deserves a number of comments. First, we should note that there
is an interesting interpretation of \eqref{opt}. While the first term inside
the maximum tends to maximize the sum of the observable joint surplus, and
hence draws the solution toward assortativeness, the second term $\mathcal{E}%
\left( \pi \right) $ is an entropic term which draws the solution toward
randomness. The trade-off between assortativeness and randomness is
expressed by the ratio $\Phi /\sigma $. If this ratio is large, the
assortative term predominates, and the solution will be close to the
assortative solution. If this ratio is small, the entropic term
predominates, and the solution will be close to the random solution. At the
same time, note that the model parameterized by $\left( \Phi ,\sigma \right)
$ is scale-invariant: if $k>0$, then the equilibrium matching distribution $%
\pi $ when the parameter is $\left( \Phi ,\sigma \right) $ is unchanged when
the parameter is $\left( k\Phi ,k\sigma \right) $. This will have important
consequences for identification, which is discussed in the next paragraph.

\bigskip

As a consequence of this result, we can deduce the equilibrium transfers and
the utilities at equilibrium. Indeed, note that combining the expression of $%
\pi $ as a function of $U$ and $a$ and equation \eqref{eqSymSchrodinger}
yields the following expression of $U$ as a function of $a$:%
\begin{equation}
U(x,x^{\prime })=\left( \Phi (x,x^{\prime })+a(x)-a(x^{\prime })\right) /2.
\label{eqUt}
\end{equation}%
which is the systematic part of utility that an individual of type $x$
obtains at equilibrium from a match with an individual of type $x^{\prime }$%
. It is equal to half of the joint surplus, plus an adjustment $%
(a(x)-a(x^{\prime }))/2$ which reflects the relative bargaining powers of $x$
and $x^{\prime }$. These bargaining powers depend on the relative scarcity
of their types; indeed, $a\left( x\right) $ is to be interpreted as the
Lagrange multiplier of the scarcity constraint which imposes that $\pi
\left( .,x\right) $ should sum to $f\left( x\right) $. Hence, the
equilibrium transfer $\tau \left( x,x^{\prime }\right) $ from $x$ to $%
x^{\prime }$ is given by%
\begin{equation}
\tau \left( x,x^{\prime }\right) =\left( \alpha (x^{\prime },x)-\alpha
\left( x,x^{\prime }\right) +a(x)-a(x^{\prime })\right) /2.  \label{tau}
\end{equation}

\bigskip

Next, note that an interesting feature of Theorem A is that, while it
characterizes equilibrium in the same-sex marriage problem, it highlights
at the same time the equivalence with the different-sex marriage problem:
indeed, as argued in DG, Theorem 1, the equilibrium matching in the
different-sex marriage problem is given by the same expression as \eqref{opt}%
, with the only difference that $\mathcal{M(}f,f\mathcal{)}$ is replaced by $%
\mathcal{M(}f,g\mathcal{)}$, where $f$ and $g$ are respectively the
distribution of men and women's characteristics.

\bigskip

We will use this characterization of the equilibrium matching as the
solution of an optimization problem in order to estimate the joint surplus $%
\Phi $ based on the observation of the matching density $\pi $. As it is
classical in the literature on the estimation of matching models with
transferable utility, the primitive object of our investigations will be the
joint surplus $\Phi $ rather than the individual pre-transfer surplus $%
\alpha $; indeed, without observations on the transfers, there is no hope to
identify $\alpha $: if we estimate that there is a high level of joint
surplus in the $\left( x,x^{\prime }\right) $ relationship, we will not be
able to determine if this is due to the fact that \textquotedblleft $x$
likes $x^{\prime }$\textquotedblright\ or \textquotedblleft $x^{\prime }$
likes $x$\textquotedblright . We will only be able to estimate that there is
a high affinity between $x$ and $x^{\prime }$.

\section{Estimation\label{sec:estimation}}


\subsection{Estimation of the affinity matrix}

Following DG, we assume a quadratic parametrization of the surplus function $%
\Phi $ to focus on a limited number of parameters which could characterize
the matching patterns. We parametrize $\Phi $ by an \emph{affinity matrix} $%
A $ so that
\begin{equation*}
\Phi _{A}(x,y)=x^{\prime }Ay=\sum_{ij}A_{ij}x^{i}y^{j}
\end{equation*}%
where $A$ has to be symmetric ($A_{ij}=A_{ji}$) in order for $\Phi $ to
satisfy the symmetry requirement. Then the coefficients of the affinity
matrix are given by $A_{ij}=\partial ^{2}\Phi (x,y)/\partial x^{i}\partial
y^{j}$ at any value $\left( x,y\right) $. Matrix $A$ has a straightforward
interpretation: $A_{ij}$ is the marginal increase (or decrease, according to
the sign) in the joint surplus resulting from a one-unit increase in the
attribute $i$ for the first partner, in conjunction with a one-unit increase
in the attribute $j$ for the second. Hence, this approach is arguably the
most straightforward way to model pairwise positive or negative
complementarities for any pair of characteristics. It does, however, not
preclude nonlinear functions of the $x_{i}$'s and the $y_{j}$'s, which can
always be appended to $x$ and $y$.

Recall equation \eqref{opt}, the optimal matching $\pi $ maximizes the
social gain%
\begin{equation}
\mathcal{W}(A)=\underset{\pi \in \mathcal{M(}f,f\mathcal{)}}{\text{max}}%
\mathbb{E}_{\pi }\left[ x^{\prime }Ay\right] -\sigma \mathbb{E}_{\pi }\left[
\ln \pi (x,y)\right]  \label{opt2}
\end{equation}%
which yields likelihood $\pi ^{A}\left( x,x^{\prime }\right) $ of
observation $\left( x,x^{\prime }\right) $, where $\pi ^{A}$ is the solution
to \eqref{opt2}. By the envelope theorem, $\partial \mathcal{W}(A)/\partial
A_{ij}=\mathbb{E}_{\pi ^{A}}\left[ x^{i}y^{j}\right] $. Hence, our empirical
strategy is to look for $\hat{A}$ satisfying
\begin{equation}
\partial \mathcal{W}(\hat{A})/\partial A_{ij}=\mathbb{E}_{\hat{\pi}}\left[
x^{i}y^{j}\right] ,  \label{matchCor}
\end{equation}%
where $\hat{\pi}$ is empirical distribution associated with the observed
matching.

\bigskip

As noted before, the model with parameters $\left( A,\sigma \right) $ is
equivalent to the model with parameters $\left( kA,k\sigma \right) $ for $%
k>0 $. Hence, a choice of scale normalization should be imposed without loss
of generality; a simple choice when a single market is considered is $\sigma
=1$, in which case the estimator $A$ is meant as the estimator of the ratio
of the affinity matrix over\ the scale parameter. The observation and
comparison of multiple markets lead to slightly different normalization
choices, which are discussed in section \ref{par:sigma}. \bigskip

If a sample of size $n$, $\left\{ \left( x_{1},y_{1}\right) ,...,\left(
x_{n},y_{n}\right) \right\} $ is observed, then $\hat{\pi}\left( x,y\right) $
is the associated empirical distribution, which places mass $1/n$ to each
observation. In DG, an estimator of $A$ is obtained by solving the following
concave optimization problem%
\begin{equation}
\min_{A\in M_{K}}\mathcal{W}(A)-\mathbb{E}_{\hat{\pi}}[%
\sum_{ij}A_{ij}x^{i}y^{j}],  \label{EstimationA}
\end{equation}%
where $M_{K}$ is the set of real $K\times K$ matrices. Indeed, the first
order conditions associated to \eqref{EstimationA} are exactly given by %
\eqref{matchCor}. However, in the present case, the symmetry of $A$ is a
requirement of the model. The population cross-covariance matrix $\mathbb{E}%
_{\pi }[x^{i}y^{j}]$ is symmetric, as $\pi $ satisfies the symmetry
restriction $\pi \left( x,y\right) =\pi \left( y,x\right) $ in the
population. Yet, in the sample, $\hat{\pi}$ has no reason to be symmetric,
as the first vector of variables $x$ typically designates the surveyed
individual, while the second vector of variables $y$ designates the partner
of the surveyed individual. Hence, the empirical matrix of covariances $%
\mathbb{E}_{\hat{\pi}}[x^{i}y^{j}]$ will only be approximately symmetric.
Thus, we symmetrize the sample by adding the symmetric households, that is,
if household $ij$ is included, meaning that individual $i$ was surveyed and
reported partner $j$, we add a symmetric household $ji$, with $j$ surveyed
and reporting partner $i$. In other words, we replace the empirical
distribution $\hat{\pi}\left( x,x^{\prime }\right) $ by its symmetric part $(%
\hat{\pi}\left( x,x^{\prime }\right) +\hat{\pi}\left( x^{\prime },x\right)
)/2$. In the sequel, $\hat{\pi}$ will denote that symmetric part. This leads
us to propose the following definition:

\begin{definition}
The estimator $\hat{A}$ of the affinity matrix is obtained by%
\begin{equation}
\hat{A}=\arg \min_{A\in M_{K}}\{\mathcal{W}(A)-\mathbb{E}_{\hat{\pi}%
}[\sum_{1\leq i,j\leq K}A_{ij}X^{i}Y^{j}]\},  \label{estimProc}
\end{equation}%
where $M_{K}$ is the set of real $K\times K$ matrices.
\end{definition}

The asymptotic behaviour of $\hat{A}$ is computed in DG, theorem~2. A word
of caution, is, however, in order. Although we have artificially doubled the
sample size, by complementing household $\left( x_{i},y_{i}\right) $ with
its mirror image $\left( y_{i},x_{i}\right) $, one should beware that the
sample size remains $n$, not $2n$. Thus, we can use directly the bipartite
estimator on the mirrored sample, with the only modification that one will
need to multiply the standard errors by a factor $\sqrt{2}$, as the
effective sample size has not doubled.

\subsection{Categorical variables}

The previous analysis can be slightly adapted to deal with the case of
categorical variables, such as race. Assume that the set of categories is
denoted $\mathcal{R=}\left\{ 1,...,r\right\} $. Assume that the individuals
are characterized by $x=\left( x^{S},x^{R}\right) $, where $x^{S}\in \mathbb{%
R}^{K}$ are socio-economic characteristics, and $x^{R}\in \mathbb{R}^{r}$ is
a vector of dummy variables $x_{i}^{R}$ ($1\leq i\leq r$) equal to 1 if
individual $x$ is of category $i\in \left\{ 1,...,r\right\} $, and zero
otherwise. We work with the following specification of the surplus%
\begin{equation}
\Phi \left( x,y\right) =\left( x^{S}\right) ^{\prime }A^{S}y^{S}+\lambda
_{R}1\left\{ x^{R}=y^{R}\right\}  \label{fullSurplus}
\end{equation}%
where $\lambda _{R}$ is \ a term that reflects assortativeness on the
categorical variable, which provides a utility increment $\lambda _{R}$ if
both partners belong to the same category. Of course, this surplus function
can be expressed multiplicatively as $\Phi \left( x,y\right) =x^{\prime }Ay$%
, where $A$ can be written blockwise as
\begin{equation}
A=%
\begin{pmatrix}
A^{S} & 0 \\
0 & \lambda _{R}I_{r}%
\end{pmatrix}
\label{formAcateg}
\end{equation}%
and hence, $A$ is obtained by running optimization problem \eqref{estimProc}
subject to constraint \eqref{formAcateg}. Note that the envelope theorem
implies that $\lambda _{R}$ is identified by the moment matching condition%
\begin{equation*}
\Pr\nolimits_{\pi }\left( x^{R}=y^{R}\right) =\Pr\nolimits_{\hat{\pi}}\left(
x^{R}=y^{R}\right)
\end{equation*}%
which states that the predicted frequency of interracial couples should
match the observed one.

\subsection{Saliency analysis\label{par:saliency}}

The rank of the affinity matrix is informative about the dimensionality of
the problem, that is, how many indices are needed to explain the sorting in
this market. To answer this question, DG introduced \emph{saliency analysis}%
, which consists of looking for successive approximations of\ the $K$%
-dimensional matching market by $p$-dimensional matching markets ($p\leq K$%
). Assume (without loss of generality as one can always rescale) that $%
var\left( X_{i}\right) =var\left( Y_{j}\right) =1$. Then saliency analysis
consists of a singular value decomposition of the affinity matrix $%
A=U^{\prime }\Lambda V$, where $U$ and $V$ are orthogonal loading matrices,
and $\Lambda $ is diagonal with positive and decreasing coefficients on the
diagonal. This idea is found in \cite{LectureHeckman2007}, who interprets
the assignment matrix as a sum of Cobb-Douglas technologies using a singular
value decomposition in order to refine bounds on wages. This allows to
introduce new indices $\tilde{x}=Ux$ and $\tilde{y}=Vy$ which are orthogonal
transforms of the former, and such that the joint surplus reflects diagonal
interactions of the new indices, i.e. $\Phi \left( x,y\right) =x^{\prime
}U^{\prime }\Lambda Vy=\tilde{x}\Lambda \tilde{y}$.

Here, we need to slightly adapt this idea to take advantage of the symmetry
of $A$ and of the requirement that the matrix of loadings $U$ and $V $
should be identical. The natural solution is the eigenvalue decomposition of
$A$, which leads to the existence of an orthogonal loading matrix $U$ and a
diagonal $\Lambda =diag\left( \lambda _{i}\right) $ with non-increasing (but
not necessarily positive) coefficients on the diagonal such that%
\begin{equation*}
A=U^{\prime }\Lambda U.
\end{equation*}%
This allows us to introduce a new vector of indices $\tilde{x}=Ux$, which
are orthogonal transforms of the previous indices. That way, the joint
surplus between individuals $x$ and $y$ is given by
\begin{equation*}
\Phi \left( x,y\right) =x^{\prime }U^{\prime }\Lambda Uy=\tilde{x}^{\prime
}\Lambda \tilde{y}=\sum_{p=1}^{K}\lambda _{p}\tilde{x}^{p}\tilde{y}^{p}
\end{equation*}%
hence this term only reflects pairwise interactions of dimension $p$ of $%
\tilde{x}$ and $\tilde{y}$, which are either complements (if $\lambda _{p}>0$%
) or substitute (if $\lambda _{p}<0$), and there are no complementarities
across different dimensions.

\bigskip

The following statement formalizes this finding:

\begin{thB}
Assume that $\mathbb{E}_{\hat{\pi}}\left[ X\right] =0$ and that $var_{\hat{%
\pi}}\left( X^{i}\right) =1$ for all $i$. Then there exists an orthogonal
loading matrix $\hat{U}$ and a diagonal $\hat{\Lambda}=diag\left( \lambda
_{i}\right) $ with non-increasing coefficients on the diagonal such that%
\begin{equation*}
\hat{A}=\hat{U}^{\prime }\hat{\Lambda}\hat{U}
\end{equation*}%
and, denoting $\tilde{x}=\hat{U}x$ and $\tilde{y}=\hat{U}y$, the estimator
of the surplus function is given by%
\begin{equation*}
\hat{\Phi}\left( x,y\right) =\tilde{x}^{\prime }\hat{\Lambda}\tilde{y}%
=\sum_{p=1}^{K}\lambda _{p}\tilde{x}^{p}\tilde{y}^{p}.
\end{equation*}
\end{thB}

\begin{proof}
Because $\hat{A}$ is symmetric, it has the following eigenvalue decomposition%
\begin{equation*}
\hat{A}=\hat{U}^{\prime }\hat{\Lambda}\hat{U}
\end{equation*}%
where $\hat{U}$ is orthogonal, and $\hat{\Lambda}=diag\left( \lambda
_{i}\right) $ is diagonal with non-increasing coefficients. Denoting $\tilde{%
x}=\hat{U}x$ and $\tilde{y}=\hat{U}y$,
\begin{equation*}
x^{\prime }Ay=x^{\prime }\hat{U}^{\prime }\hat{\Lambda}\hat{U}y=\tilde{x}%
^{\prime }\hat{\Lambda}\tilde{y}=\sum_{p=1}^{K}\lambda _{p}\tilde{x}^{p}%
\tilde{y}^{p}.
\end{equation*}
\end{proof}

In the presence of categorical variables, the presence of a block $\lambda
I_{r}$ in~(\ref{formAcateg}) reflecting assortativeness on the categorical
variable implies that the singular values of $A$ will be the singular values
of $\tilde{A}$ in addition to $\lambda $ with multiplicity $r$. Therefore,
it is recommended to perform saliency analysis simply on the upper left
block $\tilde{A}$.

\subsection{Selection issues}

The purpose of the present paper is to compare match formation across
same-sex and different-sex marriage market using the tools we developed above.
In order to do so, we need to clearly delineate what is the relevant market
in which agents match. We make the following assumption that gay men, lesbians
and heterosexuals match on segmented markets, which we formalize into:

\begin{assumption}[Exogenous selection]
\label{ass:exogenousSelection}The selection into either the same-sex or
different-sex marriage market is exogenous.
\end{assumption}

To relax this assumption, one would have to assume that all agents are
pooled together in the same market, and choose their partner's gender among
other characteristics, based on their own sexual orientation.
The marital outcome, including the gender composition of households, would
then be an equilibrium outcome resulting from a trade-off between
socio-economic complementarities and other terms reflecting interactions
between genders, and sexual orientations of the partners. We develop and
discuss this relaxed framework in appendix~\ref{app:onemarket}, which shows
that the market can be formulated as a single unipartite one where
individuals are characterized by sexual orientation and gender in addition
to other socioeconomic traits, and choose the gender of their partners among
other characteristics. The more restrictive framework provided by assumption~%
\ref{ass:exogenousSelection} can be obtained as a limiting case of the
single-market framework where the interaction between sexual orientation and
gender is predominant with respect to other characteristics, and thus the
partner's gender is fully determined by own sexual orientation and gender.

In the absence of data on sexual orientation in our database, assumption~\ref%
{ass:exogenousSelection} allows to infer sexual orientation from market
participation, and therefore it permits to perform estimation of the
affinity matrix expressing the interactions of the socioeconomic
characteristics. However, if data on socioeconomic characteristics, gender
\emph{and} sexual orientation of matched partners were available, then the
full matrix $A$ could be estimated in a straightforward manner using our
methodology, allowing to capture interactions not only between
socio-economic terms, but also between gender and sexual orientation, etc.,
as explained in appendix~\ref{app:onemarket}\footnote{%
While this may be out of reach with current large-scale datasets, it is not
unrealistic to believe that it will be possible to perform this type of
analysis in the future. The National Survey of Family Growth, for instance,
already contains detailed data on this topic, but unfortunately has no
information on same-sex partnerships. This is due to the latter being
recognized at federal level only recently.}.

As a final remark, the very fact that sexual orientation is exogenous is
itself a strong assumption, and subject to current scientific debate.
Researchers in biology, neuroscience, sexual medicine and psychology have
provided evidence on the influence of psycho-biological mechanisms on
homosexual orientation \parencite[see][]{jannini2010male,hines2011gender}. While
there is an open debate among psychologists and social scientists about the
stability of sexual behaviour\footnote{\cite{diamond2008female} have
provided the first piece of quantitative of the fluctuations of sexual
orientation among adults women using long panel data. However, psychologists
avoid talking about sexuality as a \textquotedblleft choice of
lifestyle\textquotedblright : \cite[Chapter
5]{diamond2008sexual} considers changes in sexual orientation - and in other
aspects of sexuality - as the consequence of \textquotedblleft complex
interplays among biological, environmental, psychological, and interpersonal
factors\textquotedblright.}, research on early learning during childhood
suggests that gender-typed behaviour --including sexual attraction-- is
internalized since infancy and stabilizes by late adolescence
\parencite[see
][]{hines2011gender,dillon2011sexual}.

\bigskip

\subsection{Comparison across markets\label{par:sigma}}

Affinity matrices are a useful tool to analyze marital surplus, and we would
like to use them to compare sorting patterns across different-sex and
female/male same-sex marriage markets. However, in order to achieve this, a
discussion on normalization is needed. Indeed, recall from the above
discussion that the equilibrium matching $\pi $ is the solution to%
\begin{equation*}
\mathcal{W}\left( A,\sigma \right) =\max_{\pi \in \mathcal{M}\left(
f,f\right) }\left\{ \mathbb{E}_{\pi }\left[ X^{\prime }AY\right] -\sigma
\mathbb{E}_{\pi }\left[ \ln \pi \left( X,Y\right) \right] \right\} ,
\end{equation*}%
and therefore, a matching market with affinity matrix $A$ and scaling
parameter $\sigma $ is observationally equivalent to another market with the
same distribution of types and affinity matrix $kA$ and scaling parameter $%
k\sigma $ for $k>0$. Therefore, $A$ and $\sigma $ are not jointly
identified, but only their ratio $A/\sigma $ is identified.

It is therefore useful to adopt a normalization of $\left( A,\sigma \right) $%
. For cross-market comparison purposes, the normalization $\sigma =1$
advocated in DG can be misleading, as it assumes that the standard deviation
of the heterogeneity in preferences is the same across all markets
considered. In this case, it seems more appropriate to normalize $A$ by a
factor so that the total quantity of surplus $\mathcal{W}\left( A,\sigma
\right) $ is scaled to one in each market. That is:

\begin{assumption}
\label{ass:comparison}The affinity matrix $A$ and the amount of
heterogeneity $\sigma $ are normalized so that the equality $\mathcal{W}%
\left( A,\sigma \right) =1$ holds in each market.
\end{assumption}

When considering a single market, assumption~\ref{ass:comparison} is a mere
normalization, which can be imposed without loss of generality. It implies
that the ratio of the average surplus provided by the interaction between
characteristics $i$ and $j$ of two partners in a given market, divided by
the average total surplus of a couple in that market is given by
\begin{equation*}
\frac{A_{ij}\mathbb{E}\left[ X^{i}Y^{j}\right] }{\mathcal{W}\left( A,\sigma
\right) }=A_{ij}\mathbb{E}\left[ X^{i}Y^{j}\right] ,
\end{equation*}%
and hence $A_{ij}\mathbb{E}\left[ X^{i}Y^{j}\right] $ is the share of the
average surplus explained by the interaction between characteristics $i$ and
$j$, relative to the average total surplus of a couple in that market.

On the contrary, when considering multiple markets, assumption~\ref%
{ass:comparison} is no longer an innocuous normalization. It allows for a
direct comparison of affinity matrices across markets, if one is willing to
make the restrictive assumption that the average surplus of a couple is the
same in every market. Note that, since $\mathcal{W}\left( kA,k\right) =k%
\mathcal{W}\left( A,1\right) $ holds for any scaling parameter $k\geq 0$, in
practice, we impose this normalization by first computing the estimator $%
\hat{A}$ given by~(\ref{estimProc}), and we then report $\hat{A}/\mathcal{W}(%
\hat{A},1) $.

\section{Data\label{sec:data}}

\subsection{Data on same-sex couples}

Empirical studies on same-sex marriage have traditionally needed to cope with
poor data, due to the late legal recognition of these partnerships -- still
unachieved in several countries -- and with misreporting issues, due to
social pressure on respondents. 
Social scientists have largely relied on the data collected by the US Census
Bureau for large sample analysis of same-sex unions %
\parencite{jepsen2002empirical,black2007economics,schwartz2009assortative}.
Starting from the 1990 decennial census, individuals could report themselves
as \textquotedblleft unmarried partners\textquotedblright\ within the
household, regardless of their sex, so that gay couples could be
identified. In more recent databases from the US Census Bureau, same-sex
couples are still identifiable as out-of-marriage cohabiting partners.
Indeed, although same-sex marriages have been officiated in some American
states since 2004, they were recognized at federal level only in 2013, and
currently available surveys conducted until then by the Census Bureau have
not allowed reporting marriage bonds other than different-sex unions.

Accordingly, the present work relies on the five-year Public Use Microdata
Sample (PUMS) for 2008-2012 coming from the ACS, conducted by the US Census
Bureau. We restricted our sample to the state of California, which first
legalized same-sex marriage on June 16, 2008 following a Supreme Court of
California decision, and then -- after some judicial and political
controversies that impeded the officialization of same-sex weddings from
November 5, 2008 to June 28, 2013\footnote{%
In this period, marriage licenses issued to same-sex couples held their
validity.}{} -- a decision of the U.S. Supreme Court finally accomplished
full legalization. Restricting the sample to one state allows focusing on a
marriage market with a uniform judicial framework. Moreover, in states where
same-sex marriage is recognized, estimates on the number of married same-sex
households are more reliable, and the incidence of the measurement error is
smaller \parencite{gates2010same, virgile2011measurement}.

\subsection{Descriptive statistics}

Our sample is limited to those individuals involved in a cohabiting
partnership, both married and unmarried, thus excluding singles but also
couples whose partners do not live in the same home. Each couple is
identified as a householder with his/her partner, where both share the same
ID household number. 

The main database is composed of 681,060 individuals in couples who have
completed their schooling. Because we restrict ourselves to prime age
couples (both partners 25-50 year old), the size of our sample is decreased
to 285,546 individuals. Out of them, 3,654 individuals (1.28\% of the
sample) live in same-sex couples, of which 2,034 male (0.71 \%) and 1,620
female (0.57 \%). 87.39\% of the individuals in the sample are married
heterosexuals and 11.33 \% are cohabiting heterosexuals. For estimation
purposes, after randomly selecting a subsample of different-sex couples%
\footnote{%
We randomly select 4\% of married couples and 30\% of unmarried couples.}, a
total of 9,820 couples are considered, of which 4,959 are married and 4,799
are not.

To compare different marriage markets, following \parencite{jepsen2002empirical},
the main sample is divided into four subsamples: same-sex male couples,
same-sex female couples, different-sex unmarried couples and different-sex
married couples. This repartition is based on the assumption that
individuals enter into separate markets according to their sexuality, in
line with assumption~\ref{ass:exogenousSelection}. However, another
criterion is used to differentiate two of the subgroups: married and
unmarried different-sex couples are treated as two separate subpopulations%
\footnote{%
See \cite{mourifieSiow2014} for a very interesting discussion of the
endogenous choice of the form of marital relationship.}, since empirical
evidence has reported significant differences in patterns between these two
kinds of partnership \parencite{jepsen2002empirical,schwartz2009assortative}.
Although it is impossible to know \textit{a priori} if a person is
interested in a marital union rather than in a less binding relationship,
this repartition can be of great interest and deepen the analysis.
Nevertheless, even if California represents the larger state-level ACS
sample in the US, further splitting the gay and lesbian groups
into two subgroups would imply working with potentially very small samples.
Moreover, although same-sex marriage is permitted, it has been recognized
only recently and at the end of many legal struggles, which may have
prevented a part of those same-sex couples that wished to marry from doing
so. With more data available, considering married and unmarried same-sex
couples separately would be extremely interesting, as proved by recent
research of \parencite{verbakel2014assortative} based on Dutch data.

Our study takes into consideration several variables, some related to the
labor market and some others to the general background. Non-labor market
traits include age, education and race. Age and education are treated as
continuous variables, with the latter defined as the highest schooling level
attained by the individual. Thanks to the detailed data of the ACS, the
variable has been built in order to reflect as many distinct educational
stages as possible. We consider five large racial/ethnic groups:
Non-Hispanic White, Non-Hispanic Black, Non-Hispanic Asian, Hispanic and
Others\footnote{%
American demographic institutions do not include a Hispanic category in
variables on race, furnishing a separate variable for Hispanic origins,
which is why there is some overlapping and the other categories bear the
specification ``Non-Hispanic''. The issue concerns the conceptual
differences of "race" and ``ethnicity''. See for instance \parencite%
{rodriguez2000changing} for clarifications.}.
Finally, among labor market variables, we compute and include hourly wage%
\footnote{%
The variable is computed as follows: we divide yearly wage by 52 in order to
have the average weekly wage for last year and then we divide it again by
the usual number of hours worked per week, which is available in the
dataset. The hourly wage is partly approximated because the exact number of
weeks worked in the last 12 months is not available. Note also that when
information on labor earning or number of working weeks is missing, we set
the hourly wage and the number of working weeks to 0 so that we keep all
individuals in our analysis.} and\ usual amount of hours worked per week.
Note that yearly wage is top-coded for very high values (over \$999,999).

\bigskip

Table \ref{mean} presents some descriptive statistics of our sample.
Individuals in same-sex couples are on average more educated than
individuals in opposite-sex couples. As observed by \parencite{black2007economics}%
, lesbians are much more likely to be part of the labor force than
women in different-sex couples, and also have higher wages. We observe that unmarried
different-sex couples are much younger than married couples and same-sex
couples. Unmarried heterosexual men and women are on average four year
younger than others. Cohabitation is often (but not always) a
\textquotedblleft trial\textquotedblright {} period before marriage, which
can explain this age difference\footnote{%
This would require a dynamic framework, which we don't have in our static
model. See the theoretical work of \cite{brien2006cohabitation} and %
\cite{gemici2011marriage}.}. Table \ref{share} presents the distribution of
ethnicity among couples: White individuals and Black women are
overrepresented among lesbians, while Asians and Hispanics are
under-represented in this population.

Table \ref{corr} presents correlations among traits. It shows that age
and educational attainment are much more correlated among married
different-sex couples than among unmarried and same-sex ones. Moreover,
the correlation is stronger for lesbian couples than for gay male ones.
Correlations on labor market outcomes are particularly interesting: hours
worked are negatively correlated only for married different-sex couples, a
possible clue of stronger household specialization, whereas the correlation
is positive albeit low for same-sex couples. On the other hand, wages
display a positive correlation in every market, with different-sex married
couples and male same-sex couples exhibiting the lowest correlation.

Table \ref{homratioG}, \ref{homratioL} and \ref{homratioAll} present
homogamy rates of couples with respect to race for different types of
couples. The homogamy rate is the ratio between the observed number of
couples of a certain type and the counterfactual number which should be
observed if individuals formed couples randomly. For instance, table \ref%
{homratioL} shows that lesbian couples among Black women form 10 times much
more than if they were formed randomly among the lesbian population.

\section{Results\label{sec:results}}

Homogamy rates and correlations presented in section \ref{sec:data} are
interesting measures of assortative mating and provide a good starting point
for our analysis. However, they are not sufficient to reach any conclusion
about the degree of assortativeness in the marriage market. By estimating
the parameters of the surplus function, we compare the level of
complementarity and substitutability between characteristics across
different marriage markets. This approach is consistent with Becker's model
of assortative mating, and allows us to measure the degree of
assortativeness for each combination of characteristics \textit{ceteris
paribus}. In particular, we can test whether assortativeness on observables
- notably, age, race and education - is weaker among same-sex couples, as
found by \cite{schwartz2009assortative}.

While we measure the direction and strength of interactions between
traits, we do not attempt to estimate preference and production terms
separately. Hence, we cannot tell whether marital gains differ across
markets because of differences in household production rather than pure
taste for homogamy. In particular, we cannot tell to which extent
differences in the opportunity cost of bearing and raising children affect
sorting patterns\footnote{\cite{allen2017} propose a theoretical model
which explains differences in expected matching behavior, marriage rates,
non-child-friendly activities, and fertility, based on different costs of
procreation and complementarities between marriage and children.}.
If couples wish to have genetically related children, Assisted
Reproductive Technologies imply that children inherit genetic traits from
only one out of two partners, with possibly important implications for
sorting. In our empirical analysis, we limit ourselves to the estimation of
the model on carefully chosen subsamples (e.g., childless couples) in order
to provide an intuition of where the major sources of diversity between gay
and different-sex couples lie. While not exhaustive, these robustness checks
could constitute a useful starting point for future research.

Finally, Becker's model suggests that we interpret differences in
assortativeness as a consequence of differences in marital gains, rather
than as a consequence of search dynamics (notably, geographic factors and
search frictions), segmentation into local markets along socio-economic
traits, or preferences of third parties and social pressure %
\parencite{kalmijn1998intermarriage}. In particular, gay individuals tend to
move away from their hometowns and may not be \textquotedblleft
out\textquotedblright\ at school or in the workplace %
\parencite{rosenfeld2005independence}, and this could influence the composition
of their interpersonal ties\footnote{%
For instance, online  dating among heterosexuals has been found to reduce
assortative  matching on education \parencite{hitsch2010matching}. However, as
dating apps and websites grow in popularity, thus giving access  to a larger
and larger pool of possible matches, and tend to specialize  on segmented
markets (e.g., by ethnicity or religion), we wonder if  this conclusion
still holds.}. Also in this case, some of our robustness checks can help
understand how these concurrent forces affect our results. Nonetheless, we
believe that a model explicitly accounting for such factors would be
necessary to quantify their impact on sorting patterns.\newline

We report in table \ref{MainAffinityMatrix} the estimates of the affinity
matrix for gays, lesbians, married and cohabiting heterosexuals.

\subsection{Age, education and race/ethnicity}

Our estimates of the diagonal elements of the affinity matrices are highly
positive and significant for age, education and ethnicity, which confirms
previous findings about positive assortative mating. In line with the
results by \cite{jepsen2002empirical} and \cite{schwartz2009assortative},
we find that assortativeness on age and ethnicity is comparatively weak for
male same-sex couples (0.62 for age, 0.62 for ethnicity), and progressively
stronger for female same-sex (0.79, 1.26) and unmarried different-sex
couples (1.14, 1.98), whereas married different-sex couples exhibit the
strongest complementarities (2.17, 2.49). Results on education are more
nuanced: complementarity of schooling levels is the strongest for
lesbian couples (1.19), while estimates for married same-sex (0.82) and gay couples
(0.84) are not significantly different. Finally, complementarity of schooling
levels is the lowest for unmarried heterosexuals (0.66).

Our estimates on the level of educational sorting is partly at odds with
previous findings. Empirical research on this topic mainly concluded that
assortativeness on education is weaker on both male and female same-sex
marriage markets with respect to different-sex marriage markets.
However, the social science literature\footnote{%
The main reference works about mating among gay people are listed in our
introduction. We refer to \cite{schwartz2009assortative} and %
\cite{verbakel2014assortative} who, drawing from literatures from different
social sciences, both provide a complete and updated review on this topic.}
provides a large set of explanations about why sorting patterns should
differ across different-sex and same-sex couples, and not all of them
predict that educational sorting is weaker among the latter. On the one
hand, gay men and lesbians are expected to be more inclined to \textquotedblleft
transgress\textquotedblright\ social norms and to cross socio-economic and
racial barriers when choosing their partner %
\parencite{rosenfeld2005independence,schwartz2009assortative}. Our findings
suggest that this effect might be prevalent as concerns age and ethnicity.
On the other hand, gay people are also expected to have stronger \textit{%
egalitarian preferences}. \cite{verbakel2014assortative} suggest that
similar schooling levels can lead to a more equal division of labor. Spouses
that aim to concentrate their efforts on the labor market rather than to
specialize each in a different set of skills might thus exhibit a stronger
level of assortativeness on education.

As anticipated above, childrearing is a major driver of household
specialization, and same-sex couples are less likely to have children%
\footnote{%
In our sample, among the 25-50 years-old, 14.5\% of gay men have children,
37.8\% of lesbians, 58.04\% of cohabiting different-sex couples and 83.5\%
of married different sex couples.}. Hence, we estimate the affinity matrix
using the subsample of childless couples for each of our four marriage
markets\footnote{%
We are aware that the subsample constitutes an \textquotedblleft
\textsl{artificial\textquotedblright\ marriage market, since individuals do not}
rigidly self-select into a separate market based on their preference for
having children. However, our model does not have a specification that
explicitly accounts for choices related to childbearing.} (see summary table %
\ref{Summaries} and the full tables in the online appendix). We find that,
with respect to ethnicity, both childless same-sex and childless
different-sex couples exhibit a weaker taste for homogamy compared with couples
with children of the same respective sexual orientation. Similar results hold
for sorting on age, although only differences between married different-sex
couples with and without children are significantly different. It thus seems
that individuals who plan to have children look for a more similar partner along
these two dimensions than those who do not.

When it comes to education, the picture is a bit more contrasted. As for the
previously discussed traits, one observes stronger assortativeness on
education for same-sex couples with children than for those without.
In contrast, childless different-sex couples are more assortatively matched
on education than those with children. Different-sex couples who do not plan
to have children will not benefit from large gains from specialization and may
look for a partner with similar schooling. It is interesting to note that
married childless different-sex couples are found to exhibit a higher degree of
assortativeness with respect to age, ethnicity \textit{and} education with
respect to same-sex childless couples\footnote{%
We also estimate the affinity matrix for married different-sex couples with
one and three children (see table~\ref{summaryMarried}). Our findings are in
line with what stated in the main text. The higher the number of children,
the stronger the assortativeness on age and ethnicity, and the weaker the
assortativeness on education.}.

\subsection{Labor market traits}

To describe labor market traits, we must be very cautious as these outcomes
are potentially endogenous. Since we do not observe these traits at the
moment of the match formation but possibly much later, the specialization
process at work in couples may have already begun. In particular, we expect
that this specialization effect is strong in different-sex couples, who are
more likely to have children\footnote{\parencite{antecol2013labor} and \parencite%
{jepsen2015labor} showed that to a lesser extent some household
specialization also exists within same-sex households. Moreover, %
\cite{antecol2013labor} stress that childless different-sex couples are
less specialized and thus more similar to same-sex couples.}. Raising
children takes time and many mothers leave the labor force or reduce their
working hours. Consequently, because of interrupted careers and less paid
part-time jobs, their hourly wage does not rise as much as that of their
male counterparts and we observe many associations between low-wage women
and high-wage men. This phenomenon could bias our estimates. To assess the
importance and the sign of this bias, we also perform the estimation on four
additional selected samples where the specialization effect should be
limited : 1) childless couples, 2) bi-earner couples, 3) young couples
(25-35 year old), 4) recently married couples with no children. The last
selection is only available for different-sex married couples as we observe
their wedding date; we keep couples who got married in the preceding year
and who have no children. A summary of the results is available in tables %
\ref{summaryGay}, \ref{summaryLesbian}, \ref{summaryMarried} and \ref%
{summaryUnmarried}. The full tables are available in the online appendix. We
first describe the general results obtained from the main sample.\newline

First, we measure significant positive assortativeness on hourly wages for
all types of couples, although the coefficient is higher for same-sex
couples (0.05 for gays and 0.06 for lesbians) and for different-sex
unmarried couples (0.05) than for married
different-sex couples (0.01). Furthermore, we observe negative assortative
mating on working hours for married different-sex couples (-0.04),
whereas we observe much higher and significant positive estimates for
same-sex couples (0.12 for gays and 0.20 for lesbians).
The coefficient for unmarried couples is also positive
and significant (0.09). Assortative mating on wages and working hours is
likely to be related to the presence of children. As same-sex couples are
less likely to have children, they have weaker incentives for
specialization. Unmarried couples may also have lower preferences for
children than married couples. To better understand this result, we estimate
our model on childless couples. We find that married different-sex couples
without children have a positive coefficient for both wage (0.07) and working
hours (0.12), and thus are more similar to same-sex and different-sex unmarried
couples. In this regard, our results are in line with those of
\cite{jepsen2015labor}. Similarly, the assortative mating
coefficients for wages and hours are higher -- but to a lesser extent -- for
same-sex couples without children compared to those with children.
We also perform the estimation on married couples with
only one child and married couples with three children to disentangle the
effect of the presence of children among married couples. As couples have
more children, we observe a decrease in assortative mating coefficients
for wages and hours. Similarly, as expected, the estimation on bi-earner
couples shows an increase in assortative mating on wages and
hours\footnote{%
In our analysis, bi-earner couples are couples whose both members declare
positive wage and number of working weeks.}. In the case of young same-sex
couples the comparison leads to less clear-cut conclusions due to small
sample size, while positive assortative mating on labor market traits is
stronger when comparing young different-sex couples with those in our main
sample.

The cross-estimate between the wage of one partner and working hours of the
other partner is also very interesting to analyze. We find negative
assortative mating on wages and working hours. The estimate is highly
negative for lesbians (-0.19) and different-sex married couples
(-0.09 for the interaction between wife's wage and husband's hours, -0.13
for the symmetric interaction). It is also negative and significant for gays
(-0.07). Hence, for heterosexual married couples and homosexual couples, the match gain increases when one partner
increases his/her wage and the other decreases his/her working hours. This
result is robust to the presence of children, to the age of couples and to
the bi-earner sample.

\subsection{Other cross-interactions and symmetry\label{par:testSymmetry}}

Other significant positive cross-effects have been found for some
off-diagonal elements of the affinity matrix. The parameter capturing the
interaction between wage and education is persistently high and positive.
This might suggest that higher wage individuals have a preference for more
educated partners, keeping constant their wage and all other
characteristics. Affinity between these two variables is relatively weaker
for gay men (0.13) and lesbians (0.20). Complementarity between the two inputs
is stronger for different-sex couples, but the relationship is
asymmetric: estimates for married couples suggest that complementarity in
husband's wage and wife's education is stronger (0.27) than the other way
around (0.19). However, the corresponding estimates for unmarried couples go
in the opposite direction (0.21 and 0.37 respectively). The complementarity
between the two traits might be explained by the fact that high-income
individuals - independently of their educational level - may enjoy the
company of cultured partners.

Another cross-interaction that arises from the estimation is the
substitutability between age and hours worked on same-sex marriage
markets (-0.13 for gay men and -0.10 for lesbians). This interaction might be
due to household bargaining dynamics, as explained by %
\cite{oreffice2011sexual}: younger partners enjoy higher bargaining power
and thus can afford reducing their labor supply. Interestingly, unmarried
different-sex couples exhibit similar patterns, although the effect is
weaker.

Finally, as a last robustness check for our main results, we estimate a
bipartite matching model of the same-sex marriage market where the affinity
matrix is not required to be symmetric. In this case, we need to define two
separate subpopulations to run a bipartite estimation, and therefore we need
to define household roles. On one side of the market, we group all those
partners that are registered as \textquotedblleft
householders\textquotedblright {}, whereas on the other we group their
\textquotedblleft cohabiting partners\textquotedblright . This repartition
is highly artificial, since it implies that two gay individuals that are
householders before finding a partner can never match: in general, it seems
implausible to divide the same-sex population in two separate subgroups with
the data that we have at hand. Nonetheless, it is interesting to check -
under the strict assumption of predetermined roles - whether some asymmetry
in cross-interactions occurs. We observe that the affinity matrices for both
gay men and lesbians (respectively tables \ref{Gay4} and \ref%
{Lesbian4}) are not much different than in the unipartite case. When testing
for differences between the off-diagonal coefficients\footnote{%
We provide test results in the online appendix.}, we find that only the
cross-interaction between age and hours is significantly different between
householders and their partner in male same-sex households. The interaction
is significant only in one direction, relatively young cohabiting partners
can reduce their labor supply when matching with older householders. The
difference is non-significant for female same-sex households. We already
stated that we were not very confident in the label \textquotedblleft
householder\textquotedblright\ to define a particular role in the couple. As
additional robustness checks, we assign a particular role to each partner in
same-sex couples according to another characteristic. We test if there is a
particular role assigned to 1) the older partner, 2) the higher earner
(total income) partner. Each partner belongs to a certain population
depending to his/her status in the couple, then we estimate a bipartite
matching model of the same-sex marriage market on these two populations.
Results are presented on the online appendix. Again, they are not much
different from the unipartite case. We now detail some exceptions. When
separate populations are defined according to the relative age, we now find
a negative interaction between age of the older partner and education and
hours worked of the younger partner among gay couples.
When separate populations are defined according to the
relative total income, symmetry is respected for all coefficients for male
same-sex couples but not for female same-sex couples. Specialized roles may
appear among lesbian couples as they are more likely to have children. There
are asymmetries in the interaction between hours and education, hours and
wage and education and wage. The coefficient of interaction between
education and wage is always high and positive but it is much higher between
the education of the higher earner and wage of the lower earner than the other
way around.

\subsection{Matching on unobservables}

Thanks to assumption~\ref{ass:comparison}, we can evaluate the parameter $%
\sigma $ for each market. As anticipated in section \ref{sec:model}, this
parameter has a simple interpretation: the higher $\sigma $, the more
matching appears as random to the econometrician, or, in other words, the
higher the entropy. Since our observable characteristics are meant to
capture the main socio-economic traits, we expect that a higher $\sigma $
implies that matching is less \textquotedblleft
deterministic\textquotedblright {}: indeed, for higher $\sigma $, the
socio-economic background of an individual matters relatively less, whereas
other unobservable traits (e.g. personality or physical appearance) may
matter relatively more.

We find that entropy is higher on same-sex marriage markets (1.26 for
gay men and 1.23 for lesbians), whereas it is lower on different-sex marriage
markets (1.04 for unmarried couples and 1.00 for married). Hence, if we
interpret entropy as due to the relevance of unobservables, it seems that,
among same-sex couples, socio-economic background matters less relatively to
unobservable traits.

While we privilege an interpretation of $\sigma $ that is consistent with
Becker's frictionless matching model, it is important to recall that
differences in the search process are also captured by $\sigma $. All else
held constant, stronger search frictions should result in higher entropy %
\parencite{shimer2000assortative}. However, we are not able to disentangle the
effects of search frictions from the relevance of unobservables. This may be
problematic when comparing the same- and different-sex marriage markets
because gays constitute a relatively small part of the population. If
the frequency of meetings is increasing in the size of the pool of potential
partners, as suggested by the search-and-matching literature\footnote{%
In the search-and-matching literature, the meeting rate is usually modeled
as a constant return to scale function of vacancies of both sides of the
market, usually a Cobb-Douglas function. This assumption seems to be
supported empirically on the labor market \parencite{petrongolo2001looking}. On
the marriage market, \cite{gousse2017marriage} also use a Cobb-Douglas
matching function.}, then we should expect search frictions to be stronger
for the gay population. In order to address this issue, we estimate
the affinity matrix for a subsample of couples living in the metropolitan
areas of Los Angeles and San Francisco. We expect search frictions to be
small in densely populated urban areas, and meeting opportunities to be
comparable for homosexuals and heterosexuals. While the point-estimates of $%
\sigma $ are almost unchanged for different-sex couples (1.03 for unmarried,
1.00 for married), we observe an increase for both gay men (1.35) and lesbians
(1.32)\footnote{%
Complete tables of results obtained with the subsample of couples residing
in urban areas are available in the online appendix. Urban areas are defined
as the Metropolitan Statistical Areas of Los Angeles and San Francisco.}.
Since the estimates of $\sigma $ for same-sex couples are persistently
higher even in areas where search frictions are expected to be lower, we
conjecture that the difference in entropy can, at least in part, be
explained by unobservable traits.

\subsection{Saliency analysis\label{sec:saliencyresults}}

A way to bring further insights on the main drivers of preferences of
individuals over different characteristics is to decompose the affinity
matrix in orthogonal dimensions. As detailed in section~\ref{par:saliency},
we conduct the decomposition analysis on all variables, with a specific
treatment for race, a categorical variable. Under the parametrization~(\ref%
{fullSurplus}), we estimate the affinity matrix driving the interactions of
the non-race characteristics, and the coefficient $\lambda _{R}$ which
measures the homogamy on race. We obtain%
\begin{equation*}
\Phi \left( x,y\right) =\sum_{p=1}^{K}\lambda _{p}\tilde{x}^{p}\tilde{y}%
^{p}+\lambda _{R}1\left\{ x^{R}=y^{R}\right\}
\end{equation*}%
where $\tilde{x}^{p}$ is the $p$-th index associated to the individual with
characteristics $x$ obtained by the decomposition, \ and $x^{R}$ is the race
characteristics. Taking expectations with respect to the sample distribution
\begin{equation*}
\mathbb{E}_{\hat{\pi}}\left[ \Phi \left( X,Y\right) \right]
=\sum_{p=1}^{K}\lambda _{p}\mathbb{E}_{\hat{\pi}}\left[ \tilde{X}^{p}\tilde{Y%
}^{p}\right] +\lambda _{R}\mathbb{E}_{\hat{\pi}}\left[ 1\left\{
X^{R}=Y^{R}\right\} \right] ,
\end{equation*}%
this allows us to decompose the average surplus $\mathbb{E}_{\hat{\pi}}\left[
\Phi \left( X,Y\right) \right] $ into the sum of the surplus created by the
interaction between characteristics $p$, namely $\lambda _{p}\mathbb{E}_{%
\hat{\pi}}\left[ \tilde{X}^{p}\tilde{Y}^{p}\right] $, plus the surplus
created by homogamy on the racial characteristics $\lambda _{R}\mathbb{E}_{%
\hat{\pi}}\left[ 1\left\{ X^{R}=Y^{R}\right\} \right] $. We present such a
decomposition in the appendix in table \ref{SaliencyTable}. First, we show
that the share of the average surplus created by homogamy on the racial
characteristics reaches more than 47\% for different-sex couples, 42\% for
lesbian couples and only 33\% for gays. Then, the rest of the average
surplus can mainly be explained in two orthogonal dimensions which measure
relative attractiveness. These indices load on different characteristics of
individuals. For the different-sex and the lesbian marriage market, the first
index is almost only composed of age. It explains by itself around 36\% of
the surplus for married different-sex couples, 33\% for unmarried
different-sex couples and 29\% of the surplus for female same-sex couples.
Then the second index mostly relies on education for these markets, and
explains around 27\% of the sorting for female same-sex couples, 20\% for
unmarried different-sex couples and 16\% for different-sex married couples.
For male same-sex couples, the first index of sorting relies on education
(35\%) then comes ethnicity, then age (30\%). When we consider different-sex
couples, the indices of mutual attractiveness could differ between genders.
For married heterosexuals, the education/wage index (the second index) loads
positively twice as much on wages for men than for women, whereas there is a
penalty on working hours for women that does not appear for men.


\section{Discussion and perspectives\label{sec:conclusion}}

The contributions of the present paper are twofold. From a methodological
point of view, this paper is the first to propose a tractable empirical
equilibrium framework for the analysis of same-sex marriage. Our methodology
could be applied to many other markets (e.g. roommates, teammates,
co-workers). In addition, we apply the model in order to provide an
empirical analysis of sorting patterns in the same-sex marriage market in
California. We conduct a cross-market comparison: we analyze the
heterogeneity in preferences between same-sex and different-sex couples.
First, we find that, as concerns age and ethnicity, the different-sex
marriage market is characterized by a stronger \textquotedblleft preference for
homogamy\textquotedblright {} than the same-sex marriage market. Meanwhile, results
are more nuanced when it comes to education: while lesbians show stronger
assortativeness on education, there is no significant difference in that
dimension between gay male and married different-sex couples. Second, we discuss
the differences in complementarity and substitutability in the marriage
surplus function as defined in Becker's theory of the family. Our findings
suggest that labor market traits are complementary only for same-sex and
unmarried cohabiting different-sex couples. The presence of children seems to be a
central driver of these contrasted findings. These results indicate that the
traditional concept of marriage gain based on specialization within the
couple is still relevant today, although it applies mainly to couples with
children.


Families and household arrangements are evolving quickly and we need to
understand the underlying forces of these changes. The need for effective
analytical frameworks to study and describe new forms of families has
recently emerged in the economic literature, mainly as concerns same-sex
couples \parencite{black2007economics,oreffice2011sexual} and cohabiting
partners \parencite{stevenson2007marriage,gemici2011marriage}. Let us briefly
discuss both. In this paper, we found structural differences between three
separate subpopulations divided according to sexual preferences. However,
can we state with certainty that these markets are mutually exclusive? In
fact, individuals may endogenously choose into which market they are willing
to match. The appendix explores a theoretical framework to move beyond the
exogenous selection hypothesis. While the lack of individual-level data on
sexual orientation does not allow at this stage to go beyond a theoretical
model, we trust that, with the consolidation of the same-sex marriage and
the availability of more and more accurate data, it will soon be possible to
expand our understanding on these questions. Second, cohabitation is another
developing phenomenon %
\parencite{schwartz2009assortative,gemici2011marriage,verbakel2014assortative}
and is associated with a lower degree of specialization and a lower degree
of positive assortative mating. A promising area of research would be to
understand the preferences for marriage or cohabitation jointly with sorting
preferences. \cite{mourifieSiow2014} set a first model in that direction
for different-sex couples, which could be adapted to same-sex couples using
the techniques put forward in the present paper; see also the empirical
analysis of \parencite{verbakel2014assortative}, and \cite{alden2015effect} for
the marital and fertility decisions of same-sex households.

Another topic of interest is the effect of new forms of families on the
traditional ones over time. Opponents of same-sex marriage have voiced the fear that it
will cause the marriage institution to lose its value and favor alternative
forms of families, typically more flexible/less stable, such as
cohabitation. For now, researchers have found no effect of same-sex marriage
on the number of different-sex marriages or on the number of divorces %
\parencite{trandafir2014effect,trandafir2014effect1}. However, we wonder whether
the legal recognition of same-sex marriage could someway impact the
preferences observed on the different markets. What changes should we expect
in the behavior of heterosexuals? And could it be that same-sex couples
become more homogamous as same-sex marriage is institutionalized?

Finally, in Becker's theory, a rationale for marriage is the home production
complementarities between men and women skills. However, the traditional
gains from marriage have diminished for two main reasons. First, the
progress in home technology has decreased the value of domestic production;
second, as women took control over their fertility and have become more and
more educated, their opportunity cost of staying at home has increased %
\parencite{stevenson2007marriage, greenwood2016technology}. Despite the decrease
in the gains to traditional marriage, the institution of marriage has not
disappeared. On the contrary, there has been a high demand for same-sex
legal marriage in many developed countries. \parencite{stevenson2007marriage}
argue that individuals now look for a mate with whom they \textquotedblleft
share passions\textquotedblright\ and the new rationale for marriage is now
\textquotedblleft consumption complementarities\textquotedblright\ instead
of \textquotedblleft production complementarities\textquotedblright . It is
also possible that the act of marriage itself is still considered as
intrinsically valuable for cultural and social reasons. In any case, this
evolution may lead to even higher correlation of traits. Time will tell how
these changes will impact macroeconomic outcomes, life quality and social
distance among individuals.

\bigskip

{\small $^{\flat }$}\textit{\ Economics Department, Sciences Po, Paris.}%
\newline
\textit{Address: Sciences Po, Department of Economics, 27 rue
Saint-Guillaume, 75007 Paris, France. Email: edoardo.ciscato@sciencespo.fr.}

\textit{$^{\dag }$ New York University, Department of Economics (FAS) and Department of Mathematics (CIMS), and}%
\newline
\textit{Toulouse School of Economics, Fondation Jean-Jacques Laffont. }%
\newline
\textit{Address: NYU\ Economics, 19W 4th Street, New York, NY 10012, USA.
Email: ag133@nyu.edu.} 

{\small $^{\S }$}\textit{\ Economics Department, Laval University.}\newline
\textit{Address: Universit\'{e} Laval, D\'{e}partement d'Economie, Qu\'{e}%
bec, (QC) G1V 0A6, Canada. Email: marion.gousse@ecn.ulaval.ca.} 

\printbibliography

\bigskip

\appendix

\section{Pooled matching market with sexual orientation\label{app:onemarket}}

In this appendix, we consider a model with endogenous selection of the
partner's gender. In this model, gender and sexual orientation are
observable characteristics among others. This model therefore assumes that
all the individuals are pooled into one market, and that the partner's
gender is endogenously chosen and determined based on market
characteristics, and in particular, it is subject to trade-offs with other
variables. While we include this model for completeness, we regard it as a
theoretical construction in the absence of matched data including a measure
of sexual of orientation, and we do not present it in the main paper. The
model that we use in the main body of the paper operates under the limiting
assumption that sexual orientation fully determines the partner's gender
(assumption \ref{ass:exogenousSelection}), which is why we assume in the
paper that there are three completely segmented markets: lesbian, gay, and
heterosexual.

\bigskip

In reality, segmentation may not be perfect, and some people might face
trade-offs between matching with someone of their preferred gender rather
than an attractive person of the less preferred gender. In addition, some
people might be equally attracted by both genders (i.e., bisexual).
Assumption \ref{ass:exogenousSelection} implies that we disregard these
trade-offs.

\bigskip

Consider a model where individuals are represented by: (i) their economic
characteristics $x_{e}$; (ii) their gender $x_{g}$, which is a dummy
variable equal to $0$ if male, $1$ if female (this could be extended to a
continuous gender spectrum in which case $x_{g}$ may vary continuously
between 0 and 1); and (iii) a measure of sexual orientation $x_{o}$ which is
set so that $x_{o}=1$ if the individual is maximally interested in women,
and $x_{o}=0$ if maximally interested in men\footnote{%
Sexual orientation can be measured for instance by the means of the Kinsey
scale \parencite[see ][ for a review]{sell1997defining}, which is a number
between 0 (exclusively heterosexual) and 6 (exclusively homosexual). If $k$
is the value of the Kinsey scale, $x_{o}$ will be set to $k/6$ if the
individual is a woman, and to $1-k/6$ if the individual is a man, which is
summarized in the formula $x_{o}:=x_{g}\left( k/6\right) +\left(
1-x_{g}\right) \left( 1-k/6\right) $.}. In this case, letting $x=\left(
x_{e},x_{g},x_{o}\right) $, the affinity model will result in an affinity
matrix written blockwise as%
\begin{equation*}
A=%
\begin{pmatrix}
A^{ee} & A^{eg} & A^{eo} \\
\ast & A^{gg} & A^{go} \\
\ast & \ast & A^{oo}%
\end{pmatrix}%
\end{equation*}%
where the stars denote terms that are omitted due to the symmetry of $A$.
Several components of $A$ are especially interesting:

\begin{itemize}
\item $A^{ee}$ is the classical\ affinity matrix between the socio-economic
and demographic characteristics of the partners, and measures the pairwise
assortativeness on education, age, income, race etc. Positive entries of $%
A^{ee}$ denote complementarity between these characteristics.

\item $A^{go}$ denotes the affinity between sexual orientation and gender,
which is expected to be positive, by the very definition of $x_{o}$.

\item $A^{gg}$ denotes the utility penalization of same-sex couples with
respect to different-sex ones, which we expect to be negative, in particular
due to the relatively higher cost of bearing children for these households,
and perhaps also due to social pressure in traditional societies.
\end{itemize}

In this model, there is a trade-off in surplus between a term that reflects
homogamy on socio-economic characteristics (whose strength is determined by $%
A^{ee}$), a term that reflects sexual orientation (whose strength is
determined by $A^{go}$), and a term that reflects in particular the higher
cost of bearing children (whose strength is determined by $A^{gg}$).

Note that, in the limit where the affinity term $A^{go}$ between gender and
sexual preference is very strong ($A^{go}\rightarrow +\infty $) , we get a
full segmentation of the markets, where the partner's choice is fully
determined by sexual orientation. In this case, we get a sequential choice
model where the agents choose in a first stage which market (same-sex or
different-sex) to enter, and then choose the remaining partner's
characteristics. Because we do not observe sexual orientation, we decided to
adopt this limiting case as our framework, thus making the assumption that $%
A^{go}$\ is very large with respect to the other terms. 

\clearpage
\section{Descriptive statistics}

\begin{table}[]
\caption{Sample means (25-50 year old)}
\label{mean}\centering
\resizebox{0.8\textwidth}{!}{
\begin{tabular}{lcccccc}
\multicolumn{1}{c}{\textbf{Type of couples}} & \textbf{Age} & \textbf{Education} & \textbf{Wage} & \textbf{Hours} & \textbf{Sample size} &
Share \\ \hline
\textit{Married Heterosexuals} & \multicolumn{5}{c}{} & 87.39\% \\
Men & 40.22 & 12.33 & 31.34 & 43.60 & 124,772 &  \\
Women & 38.37 & 12.47 & 22.84 & 36.16 & 124,772 &  \\
\textit{Unmarried Heterosexuals} & \multicolumn{5}{c}{} & 11.33\% \\
Men & 36.31 & 11.17 & 19.79 & 41.33 & 16,174 &  \\
Women & 34.84 & 11.55 & 18.45 & 38.30 & 16,174 &  \\
\textit{Homosexuals} & \multicolumn{6}{c}{} \\
Men & 40.00 & 13.93 & 35.19 & 42.71 & 2,034 & 0.71\% \\
Women & 39.35 & 13.78 & 28.44 & 40.74 & 1,620 & 0.57\% \\ \hline
\end{tabular}}
\end{table}

\begin{table}[]
\caption{Distribution of race by sexual orientation (25-50 year old) }
\label{share}\centering
\resizebox{0.4\textwidth}{!}{
\begin{tabular}{l|ccc|c}
Ethnic & Heterosexual & Gay & Lesbian & All \\ \hline
White & 42.7 & 67.3 & 63.5 & 43.0 \\
Black & 2.9 & 2.5 & 5.2 & 2.9 \\
Others & 0.6 & 0.8 & 1.2 & 0.6 \\
Asian & 16.7 & 8.8 & 5.8 & 16.6 \\
Hispanic & 37.1 & 20.7 & 24.3 & 36.9 \\ \hline
\textbf{Total} & 100.0 & 100.0 & 100.0 & 100.0 \\ \hline
\end{tabular}}
\end{table}

\begin{table}[]
\caption{Couples' Pearson correlation coefficients}
\label{corr}\centering
\resizebox{0.6\textwidth}{!}{
\begin{tabular}{lcccc}
\multicolumn{1}{c}{\textbf{Type of couples}} & \textbf{Age} & \textbf{Education} & \textbf{Wage} & \textbf{Hours} \\ \hline
Heterosexual married couples & 0.76 & 0.71 & 0.16 & -0.12 \\
Heterosexual unmarried couples & 0.68 & 0.64 & 0.30 & 0.11 \\
Gay couples & 0.56 & 0.56 & 0.15 & 0.03 \\
Lesbian couples & 0.66 & 0.65 & 0.20 & 0.08 \\ \hline
\end{tabular}}
\end{table}

\begin{table}[]
\centering
\begin{subtable}{0.45\textwidth}
		\centering
		\resizebox{0.95\textwidth}{!}{
			\begin{tabular}{l|ccccc}
				& White & Black & Others & Asian & Hispanic \\ \hline
				White & 1.12 & 0.79 & 0.46 & 0.94 & 0.67 \\
				Black &   & 12.31 & 0.00 & 0,44 & 0.57 \\
				Others &   &  & 40.00 & 2,14 & 1.21 \\
				Asian &   &   &   & 3.08 & 0.33 \\
				Hispanic &   &   &   &   & 2.39 \\ \hline
			\end{tabular}}
		\caption{Gays}\label{homratioG}
	\end{subtable}
\begin{subtable}{0.45\textwidth}
		\centering
		\resizebox{0.95\textwidth}{!}{
			\begin{tabular}{l|ccccc}
				& White & Black & Others & Asian & Hispanic \\ \hline
				White & 1.28 & 0.39 & 0.74 & 0.70 & 0.47 \\
				Black &   & 10.67 & 1.00 & 0.82 & 0.53 \\
				Others &   &   & 20.00 & 0.91 & 0.87 \\
				Asian &   &   &   & 8.00 & 0.13 \\
				Hispanic &   &   &   &   & 2.69 \\ \hline
			\end{tabular}}	
		\caption{Lesbians}\label{homratioL}
	\end{subtable}
\par
\centering
\par
\begin{subtable}{0.5\textwidth}
		\centering
		\resizebox{0.9\textwidth}{!}{
			\begin{tabular}{l|ccccc}
				& \multicolumn{5}{c}{\textbf{Women}} \\
				\textbf{Men} & White & Black & Others & Asian & Hispanic \\ \hline
				White & 1.96 & 0.32 & 0.87 & 0.37 & 0.28 \\
				Black & 0.49 & 24.02 & 1.38 & 0.34 & 0.36 \\
				Others & 0.84 & 0.62 & 60.91 & 0.40 & 0.46 \\
				Asian & 0.15 & 0.08 & 0.27 & 5.08 & 0.07 \\
				Hispanic & 0.26 & 0.16 & 0.44 & 0.09 & 2.32 \\ \hline
			\end{tabular}}
		\caption{Heterosexuals}\label{homratioAll}
	\end{subtable}		
\caption{Homogamy rates (25-50 year old). The homogamy rate is the ratio
between the observed number of couples of a certain type and the
counterfactual number which should be observed if individuals formed couples
randomly.}
\end{table}

\clearpage

\section*{Preamble to appendices C, D and E}

In the next three appendices, we present our estimation results. In appendix C,
we present our estimates for the main sample. Table %
\ref{MainAffinityMatrix} presents our estimates of the affinity matrix of
each market (table \ref{Gay1} for the male same-sex marriage market, table \ref%
{Lesbian1} for the female same-sex marriage market, table \ref{Married1} for the
married different-sex marriage market and table \ref{Unmarried1} for the unmarried
different-sex market). Table \ref{SaliencyTable} presents the results of our
saliency analysis, i.e., the decomposition of the affinity matrices in
orthogonal dimensions. \newline
\indent In appendix D, table \ref{BipartiteSamesex} presents our estimates
of the affinity matrix when we perform a bipartite estimation of the
same-sex marriage market without requiring the affinity matrix to be symmetric.
In this case, we define two separate subpopulations to run a bipartite
estimation. On one side of the market, we group all those gay individuals that
are registered as ``householders", whereas on the other we group their
``cohabiting partners". Table \ref{Gay4} displays our estimates for the male
same-sex marriage market whereas \ref{Lesbian4} table presents our estimates
for the female same-sex marriagemarket.\newline
\indent Finally, in appendix E, table \ref{Summaries} presents our
estimation results on additional selected samples: 1) childless couples, 2)
bi-earner couples, 3) couples living in the metropolitan area of Los Angeles
or San Francisco, 4) young couples (25-35 year old). For different-sex
married couples, table \ref{summaryMarried} also shows our results for
couples with one child only, for couples with three children and more, and
for recently married couples with no children. In this table, we do not show
all the coefficients of the affinity matrix but only the diagonal
coefficients. Each sub-table presents the results for a particular market
and each row displays the estimates for a particular selected sub-sample of
this market.

\clearpage

\section{Main estimation results}

\begin{table}[]
\centering
\begin{subtable}{0.45\textwidth}
		\centering
		\resizebox{0.95\textwidth}{!}{
			\begin{tabular}{l|ccccc}
				& \textbf{Age} & \textbf{Educ.} & \textbf{Wage} & \textbf{Hours} &
				\textbf{Race} \\ \hline
				\textbf{Age} & \textbf{0.62} & -0.06 & -0.02 & \textbf{-0.13} & \\
				& \textit{(0.04)} & \textit{(0.06)} & \textit{(0.03)} &	\textit{(0.04)} & \\
				\textbf{Education} & & \textbf{0.84} & \textbf{0.13} & -0.07 & \\
				& &	\textit{(0.09)} & \textit{(0.06)} &	\textit{(0.06)} & \\
				\textbf{Wage} & & & \textbf{0.05} & \textbf{-0.07} & \\
				& & & \textit{(0.02)} &	\textit{(0.03)} & \\
				\textbf{Hours} & & & & \textbf{0.12} & \\
				& & & &	\textit{(0.04)}& \\
				\textbf{Race} & & & & & \textbf{0.62} \\
				& & & & & \textit{(0.06)}\\\hline
				$\mathbf{\sigma}$ & \multicolumn{5}{c}{1.26} \\ \hline
			\end{tabular}}
		\subcaption{Gays (1,017 couples)}\label{Gay1}
	\end{subtable}
\begin{subtable}{0.45\textwidth}
		\centering
		\resizebox{0.95\textwidth}{!}{		
			\begin{tabular}{l|ccccc}
				& \textbf{Age} & \textbf{Educ.} & \textbf{Wage} & \textbf{Hours} &
				\textbf{Race} \\ \hline
				\textbf{Age} & \textbf{0.79} & 0.04 & 0.05 & \textbf{-0.10} & \\
				& \textit{(0.05)} &	\textit{(0.07)} & \textit{(0.06)} &	\textit{(0.05)} & \\
				\textbf{Education} & & \textbf{1.19} & \textbf{0.20} & -0.01 & \\
				& &	\textit{(0.12)} & \textit{(0.10)} & \textit{(0.07)} & \\
				\textbf{Wage} & & & 0.06 & \textbf{-0.19} & \\
				& & & \textit{(0.04)} & \textit{(0.05)} & \\
				\textbf{Hours} & & & & \textbf{0.20} & \\
				& & & &	\textit{(0.05)} & \\
				\textbf{Race} & & & & & \textbf{1.26} \\
				& & & & & \textit{(0.07)} \\ \hline
				$\mathbf{\sigma}$ & \multicolumn{5}{c}{1.23} \\ \hline
			\end{tabular}}
		\subcaption{Lesbians (810 couples)}\label{Lesbian1}
	\end{subtable}
\par
\bigskip
\par
\begin{subtable}{0.45\textwidth}
		\centering
		\resizebox{0.95\textwidth}{!}{
			\begin{tabular}{l|ccccc}
				& \multicolumn{5}{c}{\textbf{Women}} \\
				\textbf{Men} & \textbf{Age} & \textbf{Educ.} & \textbf{Wage} & \textbf{\
					Hours} & \textbf{Race} \\ \hline
					\textbf{Age} & \textbf{2.17} & \textbf{-0.20} & -0.01 & -0.03 & \\
					& \textit{(0.05)} & \textit{(0.03)} & \textit{(0.02)} & \textit{(0.03)} & \\
					\textbf{Education} & -0.04 & \textbf{0.82} & \textbf{0.19} & -0.04 & \\
					& \textit{(0.03)} & \textit{(0.03)} & \textit{(0.03)} & \textit{(0.03)} & \\
					\textbf{Wage} & \textbf{0.09} & \textbf{0.27} & \textbf{0.01} & \textbf{-0.13} & \\
					& \textit{(0.03)} & \textit{(0.04)} & \textit{(0.00)} & \textit{(0.02)} & \\
					\textbf{Hours} & \textbf{0.06} & \textbf{0.09} & \textbf{-0.09} & \textbf{-0.04} & \\
					& \textit{(0.02)} & \textit{(0.02)} & \textit{(0.01)} & \textit{(0.02)} & \\
					\textbf{Race} & & & & & \textbf{2.49} \\
					& & & & & \textit{(0.04)} \\\hline
					$\mathbf{\sigma}$ & \multicolumn{5}{c}{1.00} \\\hline
		\end{tabular}}
		\caption{Married heterosexuals (6,228 couples)}\label{Married1}
	\end{subtable}
\begin{subtable}{0.45\textwidth}
		\centering
		\resizebox{0.95\textwidth}{!}{
			\begin{tabular}{l|ccccc}
				& \multicolumn{5}{c}{\textbf{Women}} \\
				\textbf{Men} & \textbf{Age} & \textbf{Educ.} & \textbf{Wage} & \textbf{\
					Hours} & \textbf{Race} \\ \hline
					\textbf{Age} & \textbf{1.14} & \textbf{-0.06} & 0.00 & \textbf{-0.06} & \\
					& \textit{(0.03)} & \textit{(0.02)} & \textit{(0.02)} & \textit{(0.02)} & \\
					\textbf{Education} & \textbf{-0.07} & \textbf{0.66} & \textbf{0.37} & \textbf{0.05} & \\
					& \textit{(0.02)} & \textit{(0.02)} & \textit{(0.04)} & \textit{(0.02)} & \\
					\textbf{Wage} & 0.01 & \textbf{0.21} & \textbf{0.05} & \textbf{0.06} & \\
					& \textit{(0.04)} & \textit{(0.05)} & \textit{(0.01)} & \textit{(0.03)} & \\
					\textbf{Hours} & \textbf{-0.05} & -0.02 & \textbf{0.05} & \textbf{0.09} & \\
					& \textit{(0.02)} & \textit{(0.02)} & \textit{(0.02)} & \textit{(0.02)} & \\
					\textbf{Race} & & & & & \textbf{1.98} \\
					& & & & & \textit{(0.04)} \\\hline
					$\mathbf{\sigma}$ & \multicolumn{5}{c}{1.04} \\ \hline
		\end{tabular}}
		\caption{Unmarried heterosexuals (5,645 couples)}\label{Unmarried1}
	\end{subtable}
\caption{\textsc{Affinity matrix}: The tables display estimates of the
affinity matrix $A$ obtained with a sample of couples where both partners
are aged between 25 and 50. If the entry $A_{ij}$ is positive and
significant, then trait $i$ and $j$ are found to be complements in the
marital surplus function. On the contrary, if $A_{ij}$ is negative and
significant, $i$ and $j$ are substitutes. Standard errors are in
parentheses. Boldfaced estimates are significant at the 5 percent level.}
\label{MainAffinityMatrix}
\end{table}

\begin{table}[]
	\begin{subtable}{0.6\textwidth}
		\centering
		\resizebox{0.95\textwidth}{!}{
			\centering
			\begin{tabular}{l|cc|c}
				& \textbf{I1} & \textbf{I2} & \textbf{Ethnicity} \\ \hline
				\textbf{Age} & 0.11 &  0.97  & \\
				\textbf{Education} &  0.94  &-0.15 & \\
				\textbf{Wage} & 0.27 &  -0.04  & \\
				\textbf{Hours} & -0.19 &  -0.19  & \\ \hline
				\textbf{Share of systematic surplus} & 35\% & 30\%
				& 33\% \\ \hline
			\end{tabular}}
		\caption{Gays (1,017 couples)}\label{Gay3}
	\end{subtable}
	\begin{subtable}{0.6\textwidth}
		\centering
		\resizebox{0.95\textwidth}{!}{
			\begin{tabular}{l|cc|c}
				& \textbf{I1} & \textbf{I2} & \textbf{Ethnicity} \\ \hline
				\textbf{Age} & 0.98 & -0.17  & \\
				\textbf{Education} & 0.15 & 0.97  & \\
				\textbf{Wage} & 0.09 & 0.19  & \\
				\textbf{Hours} & -0.13 & -0.03  & \\ \hline
				\textbf{Share of systematic surplus} & 29\% & 27\%
				& 42\% \\ \hline
			\end{tabular}}
		\caption{Lesbians (810 couples) }\label{Lesbian3}
	\end{subtable}
	\par
	\bigskip
	\par
	\begin{subtable}{0.7\textwidth}
		\centering
		\resizebox{0.95\textwidth}{!}{
			\begin{tabular}{l|cc|cc|c}
				& \multicolumn{2}{c|}{\textbf{I1}} & \multicolumn{2}{c|}{\textbf{I2}} & \textbf{Ethnicity} \\
				& Men & Women & Men & Women &\\ \hline
				\textbf{Age}&1.05&1.02& 0.04& 0.11 &\\
				\textbf{Education}&-0.06&-0.11& 0.86& 0.97 &  \\
				\textbf{Wage}&0.03&-0.01& 0.31& 0.18& \\
				\textbf{Hours}&0.02&-0.02& 0.09&-0.14& \\ \hline
				\textbf{Share of systematic surplus} & \multicolumn{2}{c|}{36\%} &
				\multicolumn{2}{c|}{16\%} & 	48\% \\ \hline
			\end{tabular}}
		\caption{Married heterosexuals (6,228 couples)}\label{Married3}
	\end{subtable}
	\par
	\bigskip
	\par
	\begin{subtable}{0.7\textwidth}
		\centering
		\resizebox{0.95\textwidth}{!}{
			\begin{tabular}{l|cc|cc|c}
				& \multicolumn{2}{c|}{\textbf{I1}} & \multicolumn{2}{c|}{\textbf{I2}} &
				\textbf{Ethnicity} \\
				& Men & Women & Men & Women &  \\ \hline
				\textbf{Age} & 0.96 & 0.95 &  0.12 &  0.12 & \\
				\textbf{Education} & -0.13 & -0.12 &  0.97 &  0.93 &  \\
				\textbf{Wage} & -0.02 & -0.04 &  0.30 &  0.51 & \\
				\textbf{Hours} & -0.04 & -0.05 & -0.01 &  0.08 & \\ \hline
				\textbf{Share of systematic surplus} & \multicolumn{2}{c|}{33\%} & \multicolumn{2}{c|}{20\%} & 47\% \\ \hline
			\end{tabular}}
		\caption{Unmarried heterosexuals (5,645 couples)}\label{Unmarried3}\centering
	\end{subtable}	
	\caption{\textsc{Indices of attractiveness}: {\scriptsize Each column displays the	estimates of factor loadings explaining the composition of the $p$-th index of attractiveness $\tilde{x}^p$ and the corresponding share of average
	systematic surplus $\mathbb{E}_{\hat{\protect\pi}}\left[\Phi(X,Y)\right]$
	explained by such index (see section \protect\ref{sec:saliencyresults}). For
	each market, we present the two indices that explain the largest shares of
	surplus, as well as the share of surplus explained by ethnicity. Estimates
	are obtained with a sample of couples where both partners are aged between
	25 and 50.}}
	\label{SaliencyTable}
\end{table}

\clearpage
\section{Bipartite estimation for same-sex couples: head/spouse}

\begin{table}[]
	\begin{subtable}{0.45\textwidth}
		\centering
		\resizebox{0.93\textwidth}{!}{
			\begin{tabular}{l|ccccc}
				& \multicolumn{5}{c}{\textbf{Partner}} \\
				\textbf{Head} & \textbf{Age} & \textbf{Educ.} & \textbf{Wage} & \textbf{Hours} & \textbf{Race} \\ \hline
				\textbf{Age} & \textbf{0.60} & -0.06 & -0.00 & \textbf{-0.17} & \\
				& \textit{(0.04)} & \textit{(0.05)} & \textit{(0.03)} & \textit{(0.04)} & \\
				\textbf{Education} & 0.00 & \textbf{0.80} & \textbf{0.12} & -0.02 & \\
				& \textit{(0.06)} & \textit{(0.08)} & \textit{(0.05)} & \textit{(0.06)} & \\
				\textbf{Wage} & -0.01 & \textbf{0.16} & \textbf{0.06} & \textbf{-0.07} & \\
				& \textit{(0.04)} & \textit{(0.07)} & \textit{(0.02)} & \textit{(0.04)} & \\
				\textbf{Hours} & -0.05 & -0.09 & -0.04 & \textbf{0.11} & \\
				& \textit{(0.04)} & \textit{(0.05)} & \textit{(0.03)} & \textit{(0.04)} & \textit{(0.03)} \\
				\textbf{Race} & & & & & \textbf{0.65} \\
				& & & & & \textit{(0.06)} \\\hline
				$\mathbf{\sigma}$ & \multicolumn{5}{c}{1.36} \\ \hline
		\end{tabular}}
		\caption{Gays (1,017 couples)}
		\label{Gay4}
	\end{subtable}
\begin{subtable}{0.45\textwidth}
		\centering
		\resizebox{0.93\textwidth}{!}{
			\begin{tabular}{l|ccccc}
				& \multicolumn{5}{c}{\textbf{Partner}} \\
				\textbf{Head} & \textbf{Age} & \textbf{Educ.} & \textbf{Wage} & \textbf{\ Hours} & \textbf{Race} \\ \hline
				\textbf{Age} & \textbf{0.74} & 0.02 & 0.06 & -0.05 & \\
				& \textit{(0.05)} & \textit{(0.07)} & \textit{(0.05)} & \textit{(0.04)} & \\
				\textbf{Education} & 0.05 & \textbf{1.18} & \textbf{0.18} & 0.05 & \\
				& \textit{(0.07)} & \textit{(0.12)} & \textit{(0.09)} & \textit{(0.07)} & \textit{(0.04)} \\
				\textbf{Wage} & 0.02 & \textbf{0.17} & \textbf{0.08} & \textbf{-0.24} & \\
				& \textit{(0.05)} & \textit{(0.10)} & \textit{(0.03)} & \textit{(0.05)} & \\
				\textbf{Hours} & \textbf{-0.11} & -0.07 & \textbf{-0.15} & \textbf{0.21} & \\
				& \textit{(0.04)} & \textit{(0.07)} & \textit{(0.05)} & \textit{(0.04)} & \\
				\textbf{Race} & & & & & \textbf{1.18} \\
				& & & & & \textit{(0.07)} \\\hline
				$\mathbf{\sigma}$ & \multicolumn{5}{c}{1.33} \\ \hline
			\end{tabular}}
		\caption{Lesbians (810 couples)}
		\label{Lesbian4}
	\end{subtable}
\caption{\textsc{Affinity matrix}: The tables display estimates of $A$ in a
bipartite market where one side is represented by the population of
\textquotedblleft heads of household\textquotedblright\ and the other side
by the \textquotedblleft head's partners\textquotedblright. Contrarily to
the matrices $A$ estimated in \protect\ref{Gay1} and \protect\ref{Lesbian1},
now $A$ does not need to be symmetric: symmetry tests can be found in the
online appendix. We use a sample of same-sex couples where both partners are
aged between 25 and 50. Standard errors are in parentheses. Boldfaced
estimates are significant at the 5 percent level.}
\label{BipartiteSamesex}
\end{table}

\clearpage
\section{Further robustness checks}

\begin{table}[]
	\begin{subtable}{0.45\textwidth}
		\centering
		\resizebox{0.95\textwidth}{!}{
			\begin{tabular}{l|ccccc}
					& \textbf{Age} & \textbf{Educ.} & \textbf{Wage} & \textbf{\ Hours} &
					\textbf{Race} \\ \hline
					\textbf{All} & \textbf{0.62} & \textbf{0.84} & \textbf{0.05} & \textbf{0.12} & \textbf{0.62} \\
					& \textit{(0.04)} & \textit{(0.09)} & \textit{(0.02)} & \textit{(0.04)} & \textit{(0.06)} \\ \hline
					Childless & \textbf{0.56} & \textbf{0.65} & \textbf{0.06} & \textbf{0.18} & \textbf{0.46} \\
					& \textit{(0.04)} & \textit{(0.09)} & \textit{(0.02)} & \textit{(0.04)} & \textit{(0.07)} \\
					Both working & \textbf{0.64} & \textbf{1.20} & \textbf{0.06} & \textbf{0.29} & \textbf{0.56} \\
					& \textit{(0.05)} & \textit{(0.12)} & \textit{(0.02)} & \textit{(0.09)} & \textit{(0.07)} \\			
					Urban & \textbf{0.60} & \textbf{0.94} & 0.04 & \textbf{0.16} & \textbf{0.49} \\
					& \textit{(0.05)} & \textit{(0.13)} & \textit{(0.02)} & \textit{(0.05)} & \textit{(0.08)} \\
					25-35 year old & \textbf{2.48} & \textbf{1.07} & \textbf{0.35} & 0.01 & \textbf{0.88} \\
					& \textit{(0.49)} & \textit{(0.28)} & \textit{(0.13)} & \textit{(0.13)} & \textit{(0.16)} \\ \hline
			\end{tabular}}%
		\centering\caption{Gay couples}
		\label{summaryGay}
	\end{subtable}
	\begin{subtable}{0.45\textwidth}
		\centering
		\resizebox{0.95\textwidth}{!}{
			\begin{tabular}{l|ccccc}
				& \textbf{Age} & \textbf{Educ.} & \textbf{Wage} & \textbf{\ Hours} &
				\textbf{Race} \\ \hline
				\textbf{All} & \textbf{0.79} & \textbf{1.19} & 0.06 & \textbf{0.20} & \textbf{1.26} \\
				& \textit{(0.05)} & \textit{(0.12)} & \textit{(0.04)} & \textit{(0.05)} & \textit{(0.07)} \\ \hline
				Childless & \textbf{0.76} & \textbf{0.91} & 0.11 & \textbf{0.34} & \textbf{1.13} \\
				& \textit{(0.06)} & \textit{(0.15)} & \textit{(0.06)} & \textit{(0.06)} & \textit{(0.09)} \\
				Both working & \textbf{0.85} & \textbf{1.57} & 0.07 & \textbf{0.20} & \textbf{1.15} \\
				& \textit{(0.06)} & \textit{(0.16)} & \textit{(0.04)} & \textit{(0.09)} & \textit{(0.08)} \\			
				Urban & \textbf{0.87} & \textbf{1.62} & -0.00 &  0.13 & \textbf{1.08} \\
				& \textit{(0.08)} & \textit{(0.21)} & \textit{(0.04)} & \textit{(0.07)} & \textit{(0.10)} \\
				25-35 year old & \textbf{1.34} & \textbf{1.38} & -0.18 & \textbf{0.43} & \textbf{1.33} \\
				& \textit{(0.34)} & \textit{(0.27)} & \textit{(0.31)} & \textit{(0.10)} & \textit{(0.15)} \\ \hline
		\end{tabular}}%
		\centering\caption{Lesbian couples}
		\label{summaryLesbian}
	\end{subtable}	
\par
\bigskip
\par
	\begin{subtable}{0.45\textwidth}
		\centering
		\resizebox{0.95\textwidth}{!}{
			\begin{tabular}{l|ccccc}
				& \textbf{Age} & \textbf{Educ.} & \textbf{Wage} & \textbf{\ Hours} &
				\textbf{Race} \\ \hline
				\textbf{All} & \textbf{2.17} & \textbf{0.82} & \textbf{0.01} & -0.04 & \textbf{2.49} \\
				& \textit{(0.05)} & \textit{(0.03)} & \textit{(0.00)} & \textit{(0.02)} & \textit{(0.04)} \\ \hline
				Childless & \textbf{1.88} & \textbf{1.01} & \textbf{0.07} & \textbf{0.12} & \textbf{2.10} \\
				& \textit{(0.04)} & \textit{(0.04)} & \textit{(0.01)} & \textit{(0.02)} & \textit{(0.04)} \\
				One child & \textbf{2.14} & \textbf{0.81} & \textbf{0.04} &  0.02 & \textbf{2.38} \\
				& \textit{(0.05)} & \textit{(0.03)} & \textit{(0.01)} & \textit{(0.02)} & \textit{(0.04)} \\
				Three children & \textbf{2.38} & \textbf{0.75} & \textbf{0.02} & -0.02 & \textbf{2.69} \\
				& \textit{(0.05)} & \textit{(0.02)} & \textit{(0.00)} & \textit{(0.02)} & \textit{(0.04)} \\
				Newlyweds, childless & \textbf{1.40} & \textbf{1.37} & \textbf{0.02} & \textbf{0.16} & \textbf{1.81} \\
				& \textit{(0.05)} & \textit{(0.09)} & \textit{(0.01)} & \textit{(0.03)} & \textit{(0.05)} \\
				Both working & \textbf{2.44} & \textbf{0.94} & \textbf{0.04} & \textbf{0.20} & \textbf{2.36} \\
				& \textit{(0.05)} & \textit{(0.04)} & \textit{(0.01)} & \textit{(0.04)} & \textit{(0.04)} \\			
				Urban & \textbf{2.18} & \textbf{0.86} & \textbf{0.02} & \textbf{-0.04} & \textbf{2.59} \\
				& \textit{(0.05)} & \textit{(0.03)} & \textit{(0.00)} & \textit{(0.02)} & \textit{(0.04)} \\
				25-35 year old & \textbf{6.33} & \textbf{1.42} & \textbf{0.07} & \textbf{0.06} & \textbf{3.09} \\
				& \textit{(0.17)} & \textit{(0.05)} & \textit{(0.02)} & \textit{(0.03)} & \textit{(0.05)} \\ \hline
			\end{tabular}}%
		\centering\caption{Married couples}
		\label{summaryMarried}
	\end{subtable}
	\begin{subtable}{0.45\textwidth}
		\centering
		\resizebox{0.95\textwidth}{!}{
			\begin{tabular}{l|ccccc}
				& \textbf{Age} & \textbf{Educ.} & \textbf{Wage} & \textbf{\ Hours} &
				\textbf{Race} \\ \hline
				\textbf{All} & \textbf{1.14} & \textbf{0.66} & \textbf{0.05} & \textbf{0.09} & \textbf{1.98} \\
				& \textit{(0.03)} & \textit{(0.02)} & \textit{(0.01)} & \textit{(0.02)} & \textit{(0.04)} \\ \hline
				Childless & \textbf{1.11} & \textbf{0.90} & \textbf{0.04} & \textbf{0.24} & \textbf{1.47} \\
				& \textit{(0.03)} & \textit{(0.04)} & \textit{(0.01)} & \textit{(0.02)} & \textit{(0.04)} \\
				Both working & \textbf{1.19} & \textbf{0.70} & \textbf{0.18} & \textbf{0.47} & \textbf{1.83} \\
				& \textit{(0.03)} & \textit{(0.03)} & \textit{(0.02)} & \textit{(0.04)} & \textit{(0.03)} \\			
				Urban & \textbf{1.15} & \textbf{0.61} & \textbf{0.02} & \textbf{0.08} & \textbf{2.00} \\
				& \textit{(0.03)} & \textit{(0.02)} & \textit{(0.0 )} & \textit{(0.02)} & \textit{(0.04)} \\
				25-35 year old & \textbf{3.58} & \textbf{1.08} & \textbf{0.15} & \textbf{0.19} & \textbf{2.45} \\
				& \textit{(0.11)} & \textit{(0.04)} & \textit{(0.03)} & \textit{(0.03)} & \textit{(0.04)} \\ \hline
			\end{tabular}}%
		\centering\caption{Unmarried couples}
		\label{summaryUnmarried}
	\end{subtable}
	\caption{\textsc{Summary tables}: Each row displays the estimates of the
	diagonal coefficients of the affinity matrix $A$ obtained with a given
	sample. The first row (\textquotedblleft All\textquotedblright) refers to
	our benchmark results already presented in table \protect\ref%
	{MainAffinityMatrix}. The other rows refer to alternative subsamples used to
	conduct our auxiliary estimations. Complete tables with all entries of $A$
	are available in the online appendix. Standard errors are in parentheses.
	Boldfaced estimates are significant at the 5 percent level.}
	\label{Summaries}
\end{table}

\end{document}